\RequirePackage{ifpdf}
\documentclass{vgtc}

\usepackage[dvips]{graphicx}
\usepackage{subfig}
\usepackage{amssymb,amsfonts,amsmath}

\title{ContactTrees: A Technique for Studying Personal Network Data}

\author{Arnaud Sallaberry\\ \scriptsize {LIRMM, Universit\'e Paul Val\'ery Montpellier 3, France}
\and Yang-Chih Fu,  Hwai-Chung Ho\\ \scriptsize {Academia Sinica, Taiwan} 
\and Kwan-Liu Ma\thanks{Contact email: ma@cs.ucdavis.edu}\\ \scriptsize {University of California, Davis, USA} }

\abstract{
Network visualization allows a quick glance at how nodes (or actors) are connected by edges (or ties). A conventional network diagram of ``contact tree" maps out a root and branches that represent the structure of nodes and edges, often without further specifying leaves or fruits that would have grown from small branches. By furnishing such a network structure with leaves and fruits, we reveal details about ``contacts" in our \textit{ContactTrees} that underline ties and relationships. Our elegant design employs a bottom-up approach that resembles a recent attempt to understand subjective well-being by means of a series of emotions~\cite{Kahneman_2004}. Such a bottom-up approach to social-network studies decomposes each tie into a series of interactions or contacts, which help deepen our understanding of the complexity embedded in a network structure. Unlike previous network visualizations, \textit{ContactTrees} can highlight how relationships form and change based upon interactions among actors, and how relationships and networks vary by contact attributes. Based on a botanical tree metaphor, the design is easy to construct and the resulting tree-like visualization can display many properties at both tie and contact levels, a key ingredient missing from conventional techniques of network visualization. We first demonstrate \textit{ContactTrees} using a dataset consisting of three waves of 3-month \textit{contact diaries} over the 2004-2012 period, then compare \textit{ContactTrees} with alternative tools and discuss how this tool can be applied to other types of datasets.
}


\begin{document}

\maketitle







\section{Introduction}

Social networks, which are composed of actors (or nodes) and their connections (or edges), have been the subject of a very dynamic field of study for decades~\cite{Wasserman_1994,Scott_2000,Watts_2007}. As major problems facing humanity in the twenty-first century are political, economic, and social in nature, many social phenomena result from interactions among people, institutions, and markets~\cite{Watts_2004,Watts_2007}. Such interactions among various actors at different levels strongly call for interpreting social facts from a relational or structural perspective. Following the rapid rise of social media, furthermore, particularly online social networking sites such as Facebook and LinkedIn, the structures of social networks have become more complicated and more difficult to understand. With the help of emergent visualization technologies, network researchers have more tools available for identifying the key properties of large sets of network data~\cite{Baur_2001,Borgatti_2002,Graphviz,Ma_2013}.

Most of the existing methods and tools consider social networks as a whole, 
relying on a global or complete network approach.  
As comprehensive as it can be, the global approach tends to leave out 
or downplay some essential aspects of social relationships.  
For example, it is less common for a study of complete networks to focus 
on age and gender distributions of social ties, which can be more easily 
resolved by a local approach.  Although both global and local approaches 
help explain the complexity and dynamics of social networks, 
egocentric network representations reveal patterns and trends that 
global representations fail to highlight \cite{DeSolaPool_1978,Kadushin_2012}.

A local approach makes it possible to represent not only the connections among actors but also the characteristics of actors and the connections among them. To highlight the overall patterns of such connections, visualization tools often rely on tree-like network diagrams that use nodes (dots or circles) to represent actors and edges (lines) to represent connections or linkages. These contact-tree diagrams, however, typically stop at the connection level and lack further details about the elements upon which a connection is built – contacts or social interactions. 
As social relations are created and maintained by \textit{interactions} 
among actors, a visualization tool that fails to capture such interactions 
cannot fully display the dynamics of social networks. 
For researchers who aim to understand how social interactions or contacts impact 
the formation and evolution of egocentric networks, previous visualization tools 
do not meet their needs. In this paper we show how we resolve this issue with 
a new visualization design.   

We present \textit{ContactTrees}, a new egocentric visualization design that helps 
assess social interactions, compare interpersonal relationships, and make hypotheses 
about patterns or trends of contacts in everyday life. 
The design is based on a botanical tree metaphor. 
The main idea is to use the features of a tree (the structure of its branches, leaves, fruits, 
colors, \textit{etc.}) to map the properties of social interactions. This design fits well 
for visualizing many properties of egocentric networks. We present an application, 
applied to a specific dataset, to highlight some trends and hypotheses. The dataset 
contains three waves of 3-month \textit{contact diaries} from 2004 to 2012 \cite{Fu_2007} (\textit{i.e.}, 
lists of one-on-one contacts for three periods of 3 months each). 

Our first contribution to social network visualization lies in the local perspective. 
While most social network visualizations adopt a global approach, we take
an egocentric approach to highlight selected aspects of social interactions. 
The previous works devoted to local approaches mainly aim to convey the distance
between the person of focus and his or her network neighbors~\cite{Yee_2001,Heer_2005,Fisher_2005}, 
without any means to characterise these neighbors. Our approach aims to help 
identify individuals' overall ties and contacts with their network 
neighbors at a glance. Comparing the properties in several \textit{ContactTrees} 
further reveals trends about these individuals' social lives. 
By looking at \textit{ContactTrees} of an individual over different time 
periods, one can also speculate about the individual's personal and 
social life stories.

The second contribution comes from displaying leaves and fruits that 
symbolize interactions, contacts, or meetings among persons. 
While nearly all network graphs consist of nodes (actors) and edges (links or ties), 
our visualization moves a step further by incorporating critical information 
about each specific contact or meeting into the network.
Extending the use of ``contacts" as the building block of a network 
structure~\cite{Fu_2013}, we perceive social networks not only in terms of 
how actors are connected, but also in how such connections vary contact 
by contact. Because a relation is essentially established by a series 
of contacts between two actors, showing the properties of such fundamental 
building blocks by leaves and fruits greatly helps us understand 
relations and networks.

The nature of our design enables one to map various properties at 
both tie and contact levels. The mapping can be selected according to the intended tasks. 
Another strength of our design lies in the botanical metaphor used in such a context. 
In addition to producing attractive representations, our design is extensible
as it is fairly easy to add new glyphs showing other aspects of the dataset. 
With these strengths, \textit{ContactTrees} should appeal to social-science 
researchers who wish to take advantage of visualization tools that readily help 
them pinpoint critical features embedded behind the multilevel data 
in social networks.

\section{Related Work} \label{sec:related_work}

In this section, we review the current state of relational data visualization, realistic botanical tree creation, and data visualizations based on botanical metaphors.

\subsection{Visualizing relations} \label{subsec:visualizing_relations}

Most approaches to visualizing relationships are based on graphs, where nodes represent persons and edges represent the relations among them \cite{Freeman_2000}. Such approaches are closely related to the domain of graph drawing, which focuses on algorithms that help embed graphs in readable ways (see \cite{DiBattista_1999,Kaufmann_2001} for an introduction). Although the sizes of most social networks generate highly cluttered drawings, researchers of information visualization have developed many techniques to simplify representations (see \cite{Herman_2000,VonLandesberger_2011} for an overview). Some of the most powerful techniques involve clustering and navigation, such as \textit{TopoLayout} \cite{Archambault_2007}, \textit{ASK-GraphView} \cite{Abello_2006}, \cite{VanHam_2004}, \cite{Muelder_2008}, \cite{Sallaberry_2013}, and edge bundling \cite{Holten_2009}, or hybrid drawing methods like \textit{NodeTrix}, where some communities of the network are displayed as matrices \cite{Henry_2007}.

Navigating through networks from local views has also been addressed. For example, methods presented in \cite{Yee_2001} and \cite{Lee_2006} are based on tree layouts allowing users to explore a network from a given node. Van Ham and Perer also proposed a large graph visualization technique \cite{VanHam_2009} based on the computation of degrees of interest, in order to guide the user during the navigation.

More specific techniques have also been proposed. Jeffrey Heer and Danah Boyd have designed and implemented a graph visualization tool for online social networks \cite{Heer_2005}. Baur \textit{et al.} propose a software that includes graph visualizations and many network analysis metrics and techniques \cite{Baur_2001}. Fisher and Dourish \cite{Fisher_2004} developed applications based on collaboration network visualization to help coordinate and manage these collaborations. \textit{ContactMap} \cite{Whittaker_2004} is an interface for visualizing groups of one's personal contacts. A similar approach has been proposed in \cite{MacLean_2011}. In this tool, users can navigate through overlapping groups of their friends. The visualization of more structured relationships like genealogies also has been addressed in \cite{MacGuffin_2005,Bezerianos_2010,Tuttle_2010}.

Another approach of network visualizations adopts the structure of a tree. Networks can sometimes be composed of strict hierarchical relationships, in which each element is subordinated to one and only one other element. Such structures can be modeled by specific graphs called trees. These trees, however, are not to be confused with our \textit{ContactTrees}. Although both are derived from the botanical metaphor, the underlying structures are quite different. Before giving a quick overview of tree visualizations, it is important to make a remark on the tree structure itself. Consider a graph in which vertices are the persons of a social network and the edges are their links. Constructing a spanning tree from a given person produces an Egocentric structure that reveals the distance between this person and other persons in the network. 

This technique is inspiring but not sufficient for our purpose. First, not only do we focus on how network members are directly connected to a focal person, we also aim to map various properties of these members. Second, in addition to such relationships and properties, we want to further distinguish the attributes of social interactions between each network member and the focal person, contact by contact.  There are three main approaches for the visualization of trees. According to the paradigm of node-link diagrams, persons are represented by small shapes and relations by lines. A good introduction to the techniques is given by two books on graph drawing \cite{DiBattista_1999,Kaufmann_2001} and the \textit{treevis} website\footnote{http://treevis.net/} \cite{Jurgensmann_2010}. Persons can also be represented by areas and relations by the positioning of these areas. This is the case of \textit{Icicle Plots} \cite{Kruskal_1983}, \textit{Information Slices} \cite{Andrews_1998} and \textit{Sunburst} \cite{Stasko_2000}. The third approach is to visualize tree elements as nested areas. Two kind of methods have been proposed to create such maps: (1) dividing the plan recursively (a detailed overview by Ben Shneiderman, and updated by Catherine Plaisant, can be found at \cite{Shneiderman_2009}) (2) positioning leaves along space-filing curves \cite{Auber_2013}.

\subsection{Visualizations based on botanical metaphors}

Following the seminal papers of Ulam \cite{Ulam_1962} and Hondal \cite{Honda_1971},  computer  modelling  of  trees  has  been  an  active area of research. The Previous Work section of \cite{Palubicki_2009} gives a good overview of previous results. These methods attempt to characterize the way  real-world botanical trees grow. Therefore, they do not reflect a structure defined a priori. Moreover, most of them incorporate random settings, which make them incompatible with our purpose: Our visualization must incorporate a pre-defined structure and be based on a deterministic algorithm to facilitate comparisons.

To our knowledge, there are few visualizations based on botanical metaphors. One of the best known is the representation of trees (\textit{i.e.},  hierarchical data  structure) as botanical trees \cite{Kleiberg_2001}. These trees are 3D, while ours are 2D. A recurrent problem of visualizations in 3D is occlusion. So a 2D approach is more suitable for comparison. Ours are also much simpler. This is because of the input structure of the data. So it becomes easier for average users to use and interpret.
Chlan and Rheingans have also proposed a method to  visualize  hierarchical  structures  as  botanical  trees \cite{Chlan_2004}. They combine this  visualization with a branch cross-sections' one that represents properties shared by groups of elements.
\textit{Notabilia} is a visualization of \textit{Wikipedia} discussions with a nearly identical design with a botanical metaphor \cite{Stefaner_2010}, which shows similarities of the sequences of deletions in \textit{Wikipedia} discussions.
Xiong and Donath proposed a flower metaphor to visualize individual's behavior for online interactions \cite{Xiong_1999}. They show how their design helps finding information and comparing individuals. Their design is simpler than our but they cannot map as many attributes as we want to.


\section{Overview: Visualizing Contacts} \label{sec:overview}

\begin{figure*}
 \makeatletter
	\def\@captype{figure}
	\makeatother
  \centering
 \subfloat[A \textit{ContactTree} that represents the ties and contacts of a 25-year-old woman. Each small branch is a tie, with male ties on the left side of the trunk and female ties on the right. The leaves represent the contacts with these ties.]{\label{fig:egoid4}\includegraphics[height=0.26\textwidth]{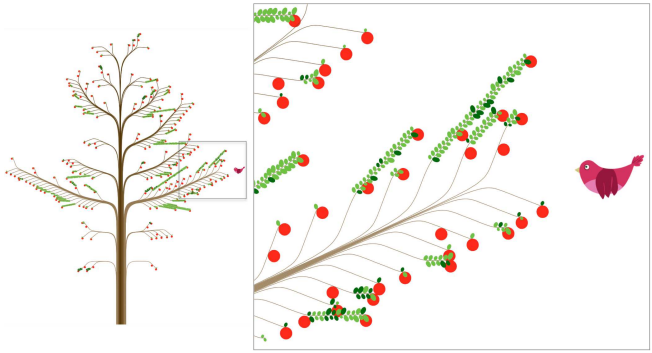}}                
  \hspace{1cm}
  \subfloat[Features of the tree on which we map properties (black labels) and an example of mapping (blue labels).]{\label{fig:overviewAbstractLabel}\includegraphics[height=0.26\textwidth]{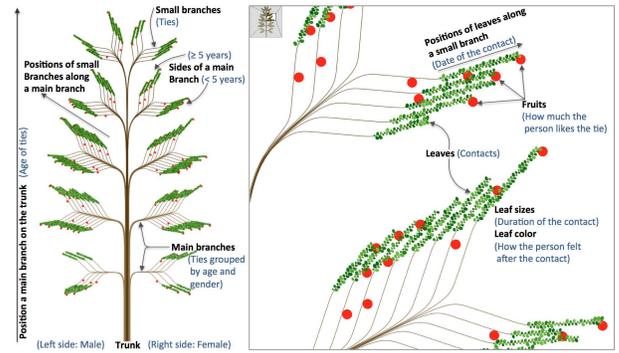}}
  \caption{Mapping properties on tree features.}
  \label{fig:fig1}  
\end{figure*}

Like other tools of tree visualization, our \textit{ContactTrees} helps visualize social relations in egocentric networks. More importantly, it also captures the properties of each specific interaction, or contact, within a relation. 
Figure \ref{fig:egoid4} shows a \textit{ContactTree} representing the ties and contacts of a 25-year-old woman (\textit{i.e.} the persons this woman met during a given period and the contacts between these persons and this woman).
Among the reasons that make us develop a botanical tree metaphor, four deserve to be highlighted :
\begin{itemize}
\item many properties can be mapped onto tree features;
\item it is easier to remember the main aspects of a tree than the main aspects of a more abstract visualization;
\item the design helps users compare properties of ties and contacts and to identify how they evolve over time;
\item an attractive design can help increase the adoption of information visualization techniques in social science.
\end{itemize} 

A 2D botanical tree structure can unambiguously hold many properties. 
For example, main branches growing from the trunk are on either the left or the right side of the trunk.
They also have a y-position on the trunk, an angle, a length, \textit{etc.}
All of these features can be used to symbolize properties.
We have selected some of them, as described in the next section.
We then present an example of mapping.

\subsection{Features of the botanical structure used to map properties}\label{subsec:overview1}

First of all, a tree can be seen as a structure composed by a trunk, which is divided into several part to form main branches,  which are further divided into smaller branches, \textit{etc.}
Starting from this observation, an intuitive approach consists of representing ties as the smallest branches (see black labels on Figure \ref{fig:overviewAbstractLabel}).
According to this, a small branch holds exclusive properties of a tie, while bigger branches represent common properties of the ties shared by the small branches growing out of them.
Finally, the aspect of the trunk represents the whole set of relationships.

The starting point of each main branch growing up from the trunk can be characterized by two parameters: its position along the trunk and the side of the trunk on which it is located (see Figure~\ref{fig:overviewAbstractLabel}). The position along the trunk is ranked on an ordinal scale, and the side of the trunk is a boolean one (either right or left). These two properties characterize the set of ties represented by the main branches.

The same approach is also valid for the main branches. The position along the main branch and the side (above/below) from where the smallest branch is starting can be used to represent properties (one ordinal and one boolean). The boolean property depicted by the side, however, is less salient than the one represented by the side of the trunk. Users tend to pay less attention to it, and thus it should be reserved for less important properties. Because the image shows that the smallest branches start from the main branches, we could also divide main branches into smaller branches holding several ties, and so on, to display more properties, but we prefer to address this task as a future work. Indeed, it is not obvious that deeper and more complex decompositions in visualization would help reveal ties and contacts more clearly, or allow end users to apply their own data to the design as easily.

In our design, each small branch represents a tie, while leaves and fruits symbolize attributes of contacts. Each leaf/fruit also holds four features on which we can map a property: color, size, position along the branch, and the side of the branch from which they are growing (see Figure \ref{fig:overviewAbstractLabel}).

Looking forward at the properties one can map on a \textit{ContactTree}, we see first that there are two boolean ones (side of the trunk, side of the main branches) and two ordinal ones (position along the trunk, position along the main branches) for grouping the ties. Then for each tie, we have two numeric properties (leaves and fruits). If they encode specific elements, two numeric properties of these elements (color, size) can be shown, as well as an ordinal property (position along the small branch) and a boolean property (side of the small branch).

\subsection{Example of mapping} \label{subsec:overview2}

Thanks to the tree features described above, many properties of ties and contacts can be mapped. The example of mapping presented here is based on \textit{contact diaries}, a major approach of data collection in the social network literature \cite{DeSolaPool_1978,Fu_2007,Lonkila_1999}. A diary normally contains inclusive information about persons with whom the diary keeper has contact within a given period of time, their relationships with the diary keeper, and the situations of each one-on-one contact. Some diaries are more specific about certain properties. For example, contact situations might include how much the diary keeper likes a contacted person, or how the diary keeper feels after each unique contact. In this paper we use an actual dataset of \textit{contact diaries}.

Blue labels in Figure \ref{fig:overviewAbstractLabel} show an example of mapping. Here, branches lying on the left (resp. right) side  of  the  trunk  are male (resp. female) ties. Positions of the main branches along the trunk represent age groups (or age cohorts) of the contacted persons: ages 0 to 9 for the  lowest main branch, ages 10 to 19 for the second-lowest branch, and so on. The tree drawn by these principles so far resembles a population pyramid in terms of a gender/age structure diagram, thus linking our design to conventional demography too. Furthermore, sides of the main branches indicate how long the diary keeper has known the contacted person (at least 5 years if a small branch is above the main branch, less than 5 years if it is below).

Each leaf is a contact between the focal person and the tie represented by the small branch from which it is growing. Leaves are ordered along a small branch by the date when the contact occurs (we do not distinguish the side of the small branch from which a leaf grows). The darker a leaf is, the better the person felt after the contact. The size of each leaf varies depending on the duration of the contact. Finally, fruits signify how much the person likes the tie (no fruit means ``not at all" or ``not much"; 1 fruit means ``somewhat"; 2 fruits mean ``very much").

Typical questions supported by our design are ``What gender is most present in a person's relationships?," ``What is the age distribution of these relationships?," ``Who are the persons one prefers?," ``How do all these properties evolve for a given person?," or ``Are there some trends among the relationships of persons having a child?" The reconfigurable interface described in the next section allows a user to select the attributes mapped into the \textit{ContactTrees} according to the objectives of her analysis or presentation. Features supported by our approach can be divided into three main categories, each of which is further divided into smaller types of features:
\begin{enumerate}
\item \textbf{Global aspect of a \textit{ContactTree}:} Balance (side of the trunk, side of the main branches), Distribution of the values (positions along the trunk and along the main branches), Outliers in terms of quantity of contacts (size of the small branches).
\item \textbf{Details of the ties of a \textit{ContactTree}:} Number of contacts (number of leaves), Qualities of the contacts (length and color of the leaves), Quality of a tie (number of fruits).
\item \textbf{Comparison of several \textit{ContactTrees}:} Trends among interpersonal relationships (tree shapes), Evolution of  individual relationships comparing two time-steps (global tree shapes refined by the local aspects of the trees).
\end{enumerate}

We will see in Case studies section how it can be useful to compare persons and discover the trends among them.

\subsection{An interface for visualizing ContactTrees}

A potential issue with our visualization lies in the burden of 
remembering visual encodings. This problem can be somewhat alleviated 
by implementing an interface that includes a legend, with which
social scientists who have used our system did not 
encounter difficulties in interpreting the \textit{ContactTrees} 
quickly. 

Which mapping to choose highly depends on the user's purpose or the task
to perform. We provide a reconfigurable interface where the user can
select mappings and explore a dataset on his own, finding the
best perspective for his tasks or comparing mappings if needed.

Our prototype has been developed as a plugin of the framework 
\textit{Tulip}\footnote{http://tulip.labri.fr/TulipDrupal/} 
\cite{auber_2003}. This framework includes a powerful data model 
and rendering system as well as many interaction techniques useful 
for our purposes. In particular, the user can zoom in/out using the
mouse wheel or selecting a rectangle with the mouse. So it is easy to access
the details of small features like fruits or leaves. When zoom in,
a small rectangle at the top left corner of the view shows the overall
map, highlighting the area one is looking at. So the main area is
devoted to the focus while the rectangle shows the context. Another 
related technique available is a fisheye lens. The user can also move 
the map with a drag'n drop, search branches to select ties sharing 
a given property, take snapshots, \textit{etc.} A video showing the
interface is available {\tt \small (https://www.youtube.com/watch?v=zxnTU7g9DiY\&feature=youtu.be)}.


\section{Algorithm} \label{sec:algorithm}

Our tree construction algorithm is based on three main steps presented hereafter.

\subsection{Step 1: Ordering ties}

\begin{figure*}
 \makeatletter
	\def\@captype{figure}
	\makeatother
  \centering
  \subfloat[Non ordered nodes.]{\label{fig:nodes_nordered}\includegraphics[width=0.4\textwidth]{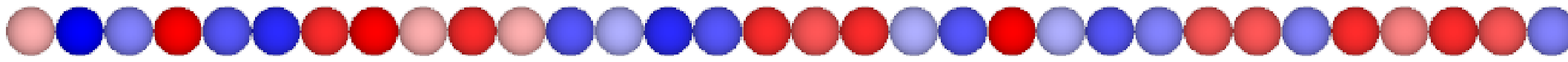}}                
  \subfloat[Ordered nodes.]{\label{fig:nodes_ordered}\includegraphics[width=0.4\textwidth]{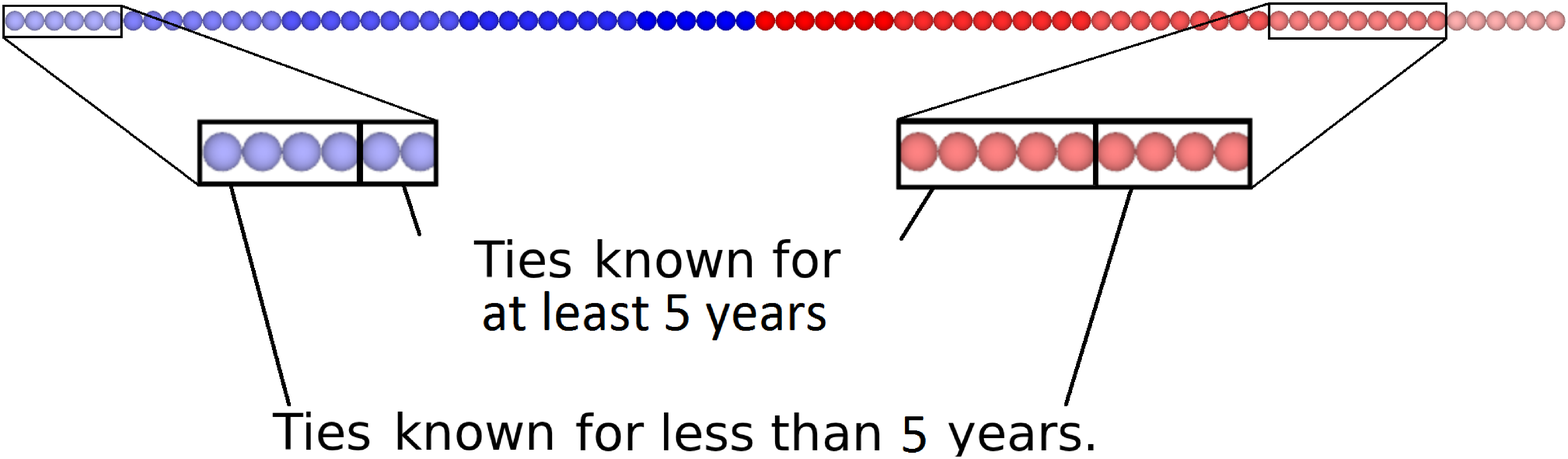}}
  \caption{Ordering of the relations according to the mapping presented in Figure \ref{fig:overviewAbstractLabel}.}
  \label{fig:nodes}
\end{figure*}

Our first step is to order the ties according to the three properties we want to take into account (see previous section). Figure \ref{fig:nodes_nordered} shows non-ordered nodes representing the ties. Blue nodes are the ties represented by branches on the left side of the tree (\textit{e.g}., males in the example of the previous section); red nodes are the ties represented by branches on the right side (\textit{e.g.}, females). Saturation depicts the properties corresponding to their positions along the trunk (\textit{e.g.}, age groups). These colors are used to help clarify our method. 

Nodes are then ordered as follows: blue nodes are first positioned on the left from the less saturated to the more saturated, red nodes are then positioned on the right from the more saturated to the less saturated. Nodes representing ties lying on the same main branch (same color and saturation, \textit{e.g.}, same gender and same age group) are ordered according to the side of the main branch from which their small branches originate (\textit{e.g.}, the number of years the person has known these ties), in increasing order for the left nodes and decreasing order for the right nodes. Figure \ref{fig:nodes_ordered} shows the result.

\subsection{Step 2: Drawing branches}

Our trees are made up of lines, with each line representing a tie. A line is at first a part of the trunk; then it becomes part of a main branch. Finally, the last segment is the small branch on which the leaves lie.

\begin{figure*}
 \makeatletter
	\def\@captype{figure}
	\makeatother
  \centering
  \subfloat[Trunk: first segment of the lines.]{\label{fig:trunc-lines}\includegraphics[height=0.25\textwidth]{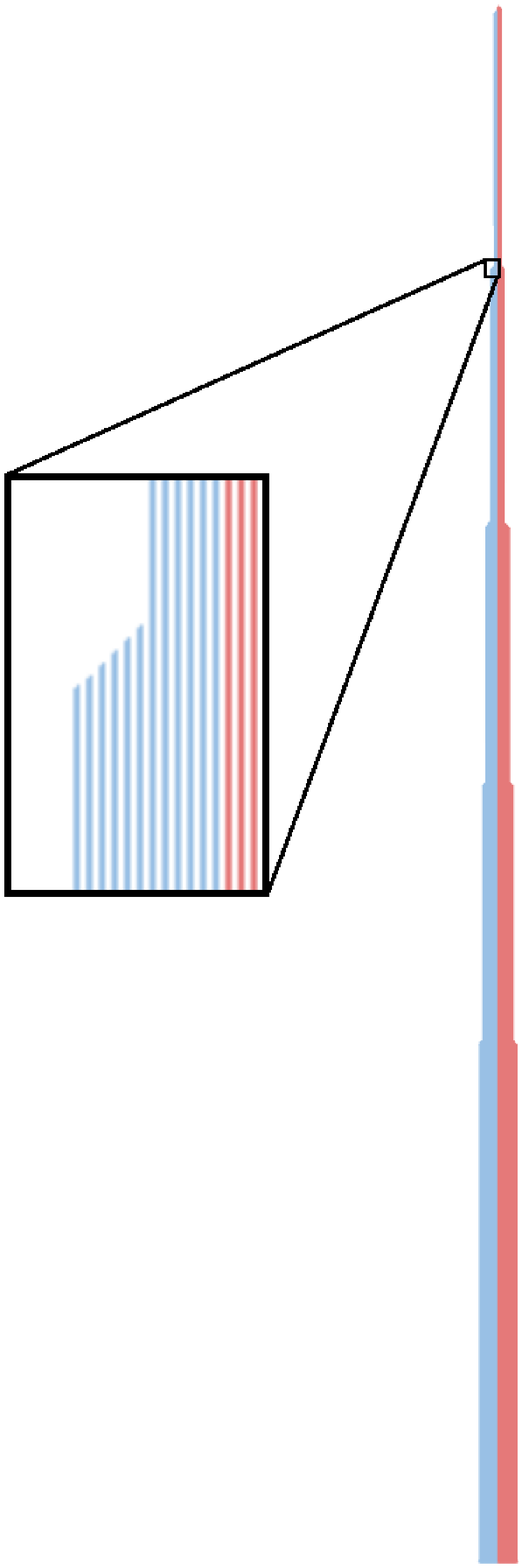}}                
  \hspace{1cm}
  \subfloat[Main branches: second and third segments of the lines.]{\label{fig:bigBranches-lines}\includegraphics[height=0.25\textwidth]{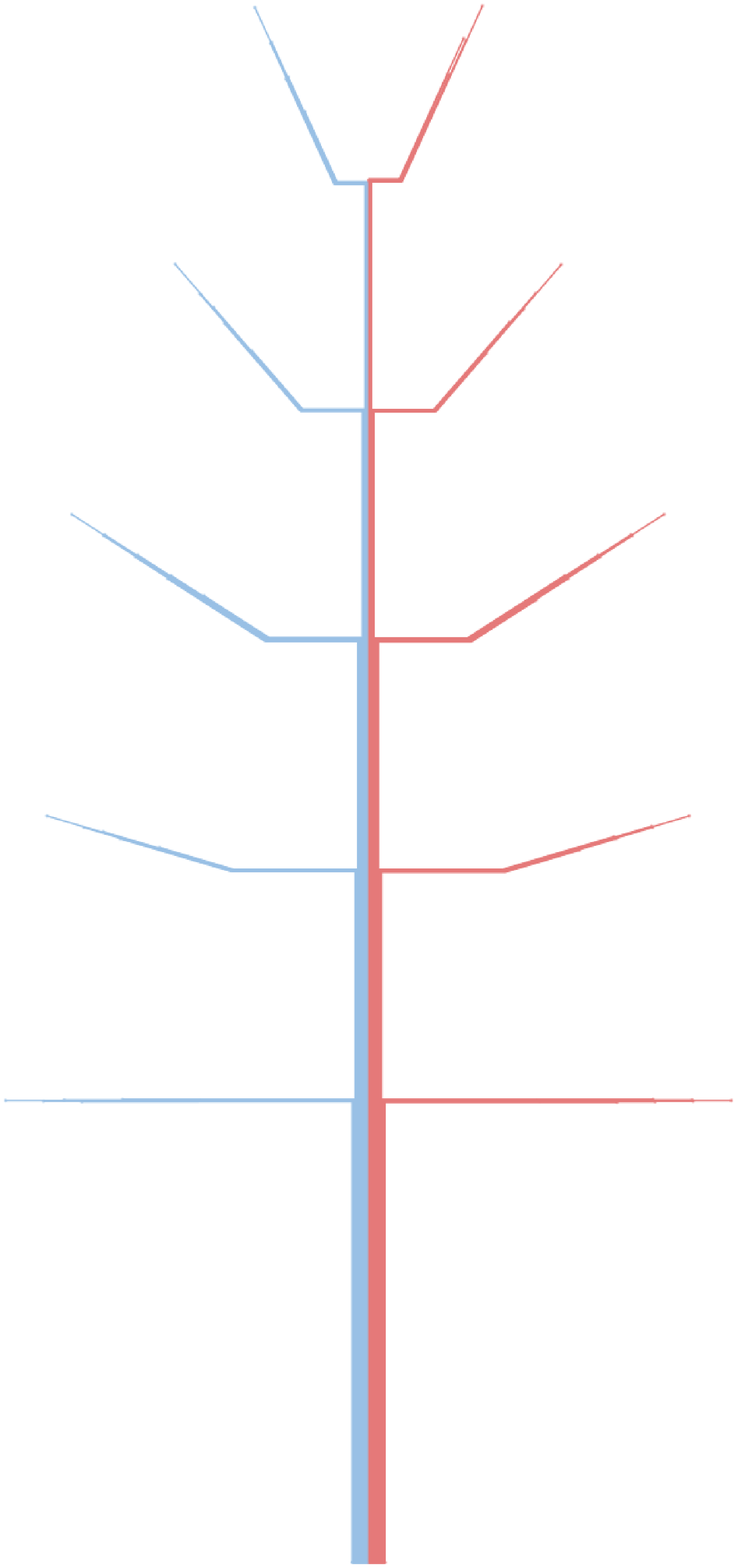}}
    \hspace{1cm}
  \subfloat[Lines end uniformly on the upper/lower side of the branch according to the number of years the person has known the ties.]{\label{fig:bigBranches-lines-zoom}\includegraphics[height=0.14\textwidth]{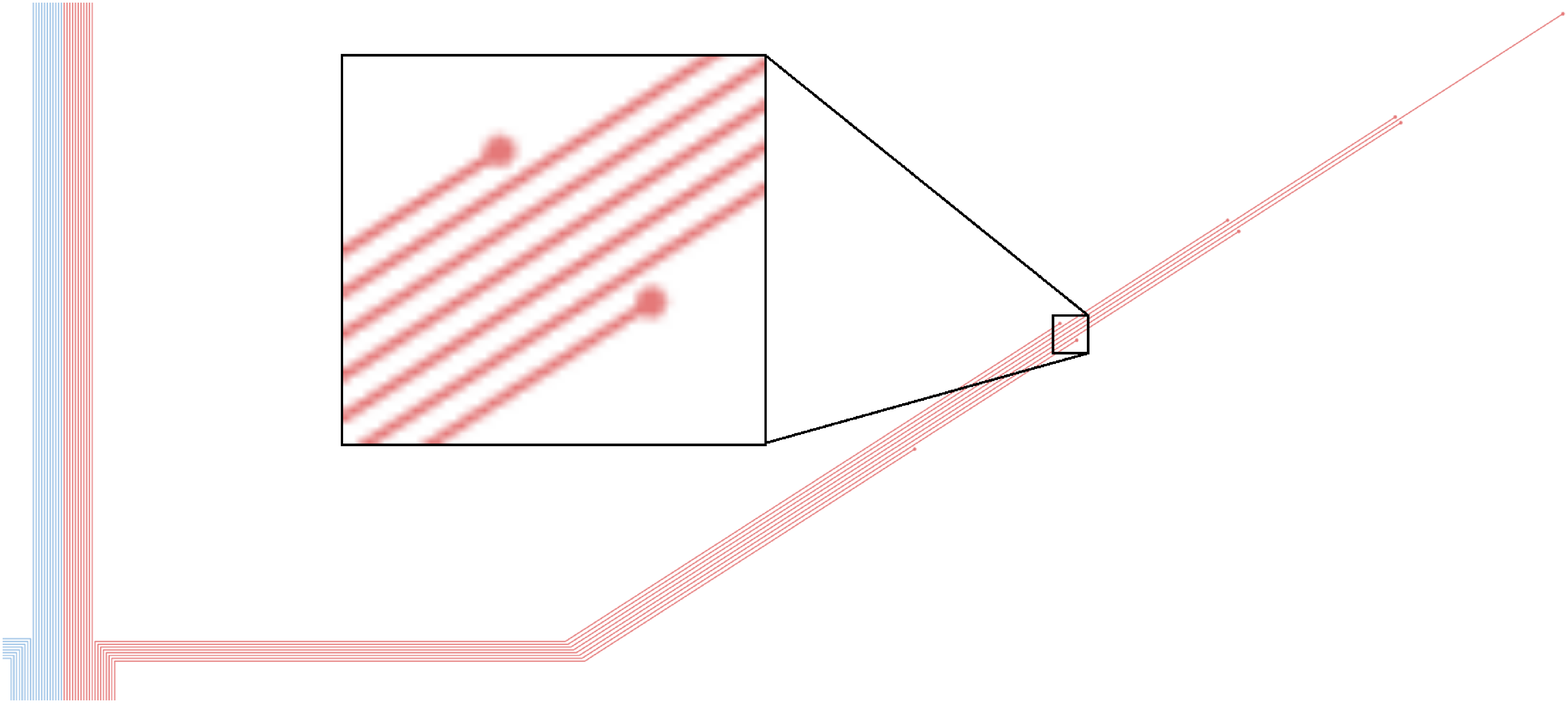}}   
  \hspace{1cm}
  \subfloat[Small branches: fourth and fifth segments of the lines]{\label{fig:smallBranches-lines}\includegraphics[height=0.28\textwidth]{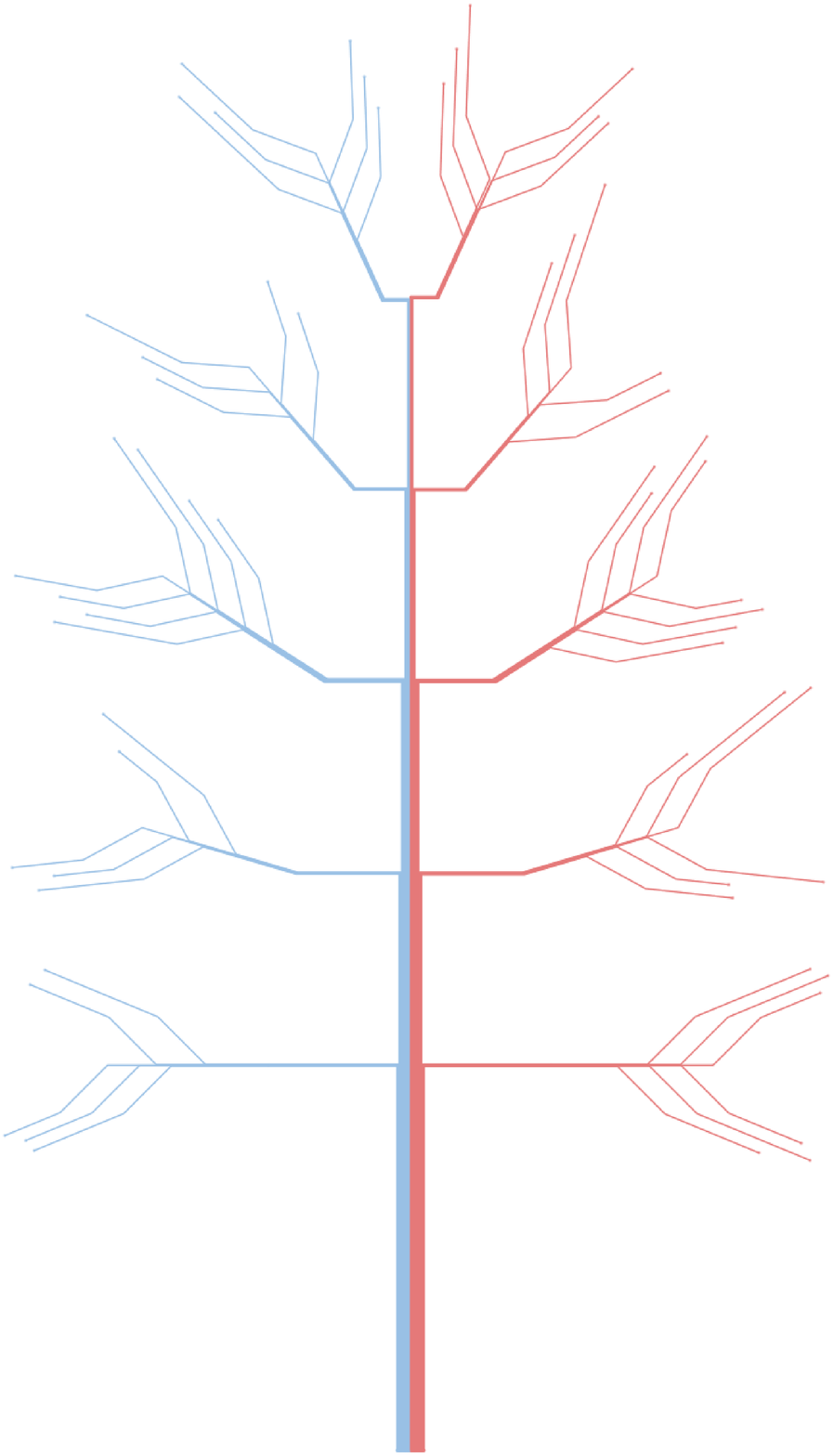}}       
  \caption{Trees are made up of lines.}
  \label{fig:trunc_bigBranches}
\end{figure*}

All lines start from the ordered nodes created in Step 1. Nodes form the basis of the trunk. The length of lines depends on the property mapped on the position of main branches along the trunk. For example, a line is longer for an older tie in the previous example (see Figure \ref{fig:trunc-lines}).

Then, second and third segments are added (see Figure \ref{fig:bigBranches-lines}). As these two new segments are positioned higher along the trunk (\textit{i.e.}, when the contacted persons are older), the second segment gets shorter and the angle between the two segments becomes sharper. These two design choices yield a triangular-shaped structure that is a more botanical tree-looking shape. The length of the third segment varies on the order computed during the previous step, and also on the side of the main branch from which we want them to grow: In the example of Figure \ref{fig:overviewAbstractLabel}, the ties known by the person for at least 5 years end uniformly on the upper side of the branch; those known less than 5 years end uniformly on the lower side (see Figure \ref{fig:bigBranches-lines-zoom}).

Finally, two more segments (the fourth and fifth)
are added to separate the lines from one another (see Figure
\ref{fig:smallBranches-lines}). 
The length of the fourth segment is a constant, as the
angle between segments. These constants do not display
information but are required to obtain a botanical 
tree-like drawing. The length of the fifth segment depends
on the number of leaves growing from the small branch.
Thus, the more contacts a tie involves, the longer the
fifth segment will be. 

\subsection{Step 3: Adding leaves and fruits}

Leaves are then added according to the side of the small branch upon which they have to lie. This side can represent a given property as mentioned in the overview section. Fruits are all displayed on the lower side of the small branch in order to fulfill the botanical metaphor. The size and color of the leaves and fruits all depend on selected properties. As an example, the darkness of leaves in Figure \ref{fig:overviewAbstractLabel} stands for how positively the person feels about a contact, and their size indicates the duration of a contact.

Lines are finally replaced by B\'ezier curves going through points uniformly distributed along them. Botanical tree-like colors are used. These two design choices make our visualization look more like a botanical tree (see Figure \ref{fig:egoid4}).

\subsection{Step 4 (optional): Adding more features}\label{subsec:more_features}

A nice aspect of our visualization is its capacity for evolution. One can easily define new features that fulfill the botanical metaphor to highlight some other aspects of a dataset. An example of this property is shown in Figure \ref{fig:egoid4}. The \textit{ContactTree} represents a person's ties and the contacts this person has made with these ties. Mapping is the same as that described in the Overview section. As we will see in the case studies section, one purpose of our visualization is to allow more intuitive and direct comparisons of contacts and ties. In this case, displaying some properties of the person, such as gender, age and marital status, should help reveal trends among relationships. Of course, such properties instead could have been displayed as text near the \textit{ContactTree}, but this option has been dismissed for two reasons. First it would not fulfill the botanical metaphor and would give to the visualization an aesthetically unpleasant aspect. Second, a more intuitive feature could be used to clearly identify gender and age groups of the focal person. These two properties are already mapped for the ties by the side of the trunk and the positions of main branches. The same principles should be used to map the properties of the focal person, too.

That is why we have introduced a bird (Figure \ref{fig:egoid4}). Its  y-position  shows  the  age  of  the person: The bird is in front of the 3rd main branch, so the person is between 20 and 29 years old, as are other ties represented by this main branch. In the same way, the bird is on the right side of the trunk. This means that the person is a woman, as are all the ties of the right part of the trunk. Finally, we use two birds when the person is married and only one (as in Figure \ref{fig:egoid4}) when he/she is not. Displaying married birds near the spouse of the person represented by the tree has also been considered. This design adds information and would be intuitive for married people, but would be misleading in the case of single persons.

The  possibility  of  adding  more  features depending
on the dataset is a particularly promising aspect of our
visualization,  which  is  only  limited  by  the designer's creativity.
While it makes the visualization more
attractive for the masses, it is also a good way to capture
more properties. As an example, different birds could be
used to highlight a handful of very special people.


\section{Case studies} \label{sec:case_studies}

The main purpose of our visualization is to help network researchers and other end users to compare both ties and contacts of several entities. Thanks to the Egocentric perspective, this comparison can involve properties of different persons, ties, and contacts. It also allows direct comparisons of how personal networks change over time. In this section, we use data from \textit{contact diaries} to demonstrate how our design helps reveal such network structures and changes.

Our case study is based on a series of \textit{contact diaries} that contain one-on-one interpersonal ties and contacts recorded by volunteers for a total of 3-9 months. This study is composed of three waves of diary keeping. In early 2004, 54 individuals in Taiwan completed their \textit{contact diaries} for 3 months. In 2008, 28 of these volunteers followed-up with the same diaries for another 3 months. In 2012, 12 of them again finished the diaries for 3 months. Each contact involves one diary keeper and a unique person who is tied to that diary keeper. A diary keeper could meet one of these persons several times during the given periods.

In the datasets, a total of 49 properties have been recorded
in or extracted from each of the 175,597 entries (contacts). 
These properties cover the
situations of each contact (\textit{e.g.}, date, duration, and how
the diary keeper felt after the contact), for each of 41,673 contacted persons (\textit{e.g.}, age, gender, and occupation),
and  the  relationship  between  each  diary  keeper  and
each contacted person (\textit{e.g.}, how many years they had
known each other, and how much the diary keeper liked
the contacted person). With the overwhelmingly rich
and complex information \textit{contact diaries} have yielded, 
it should be easy to identify patterns or trends of ties 
and contacts by using several mapping techniques from our design.

\subsection{First mapping}

The first set of examples is based on the mapping described in the Overview section.

More than just displaying data, our visualization can be used to make hypothesis about a person's social life and identify some trends among all the persons. The example presented here focuses on couples and their children.

\begin{figure*}
 \makeatletter
	\def\@captype{figure}
	\makeatother
  \centering
  \subfloat[Man]{\includegraphics[height=0.35\textwidth]{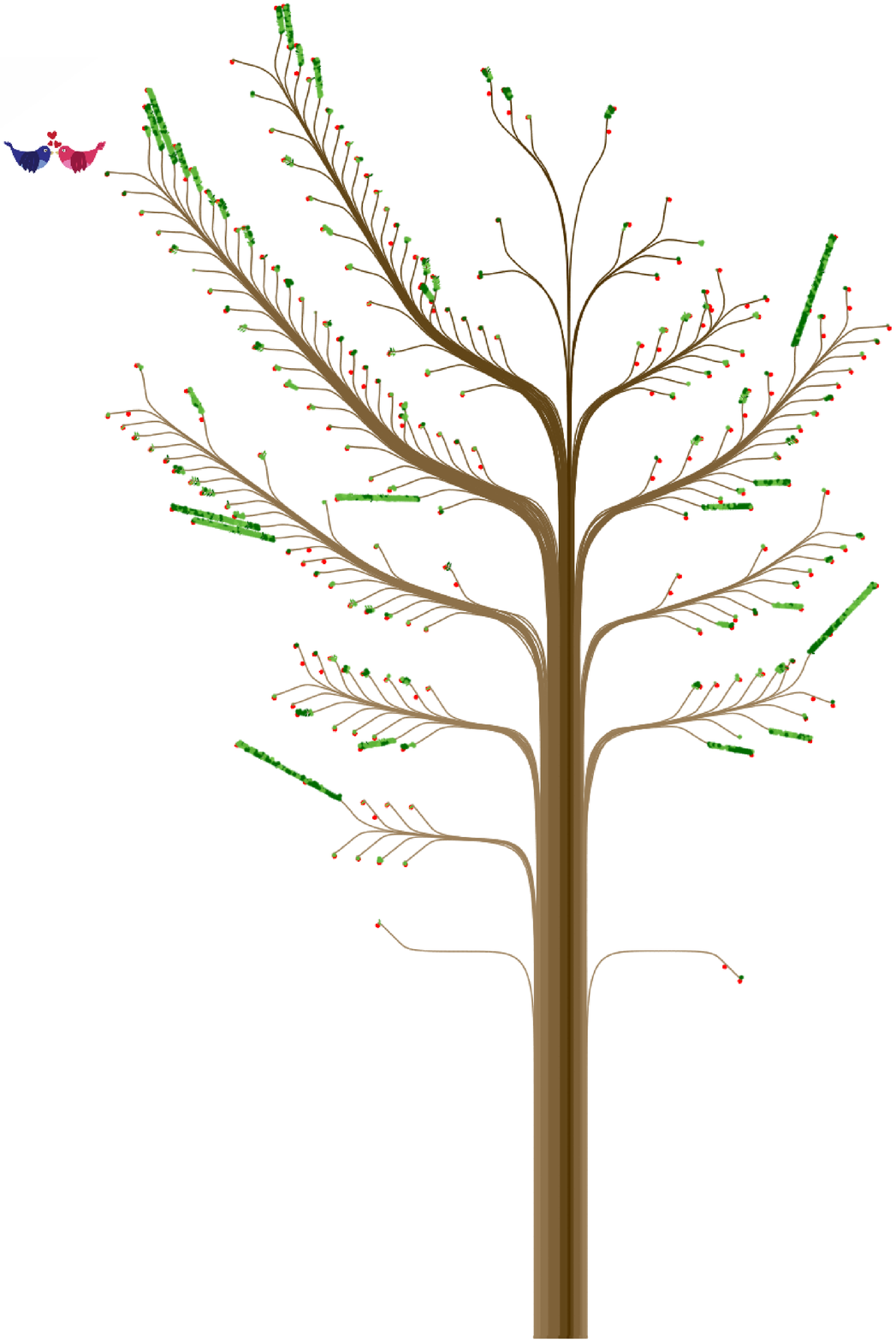}}
  \hspace{0.05\textwidth}             
  \subfloat[Woman]{\includegraphics[height=0.35\textwidth]{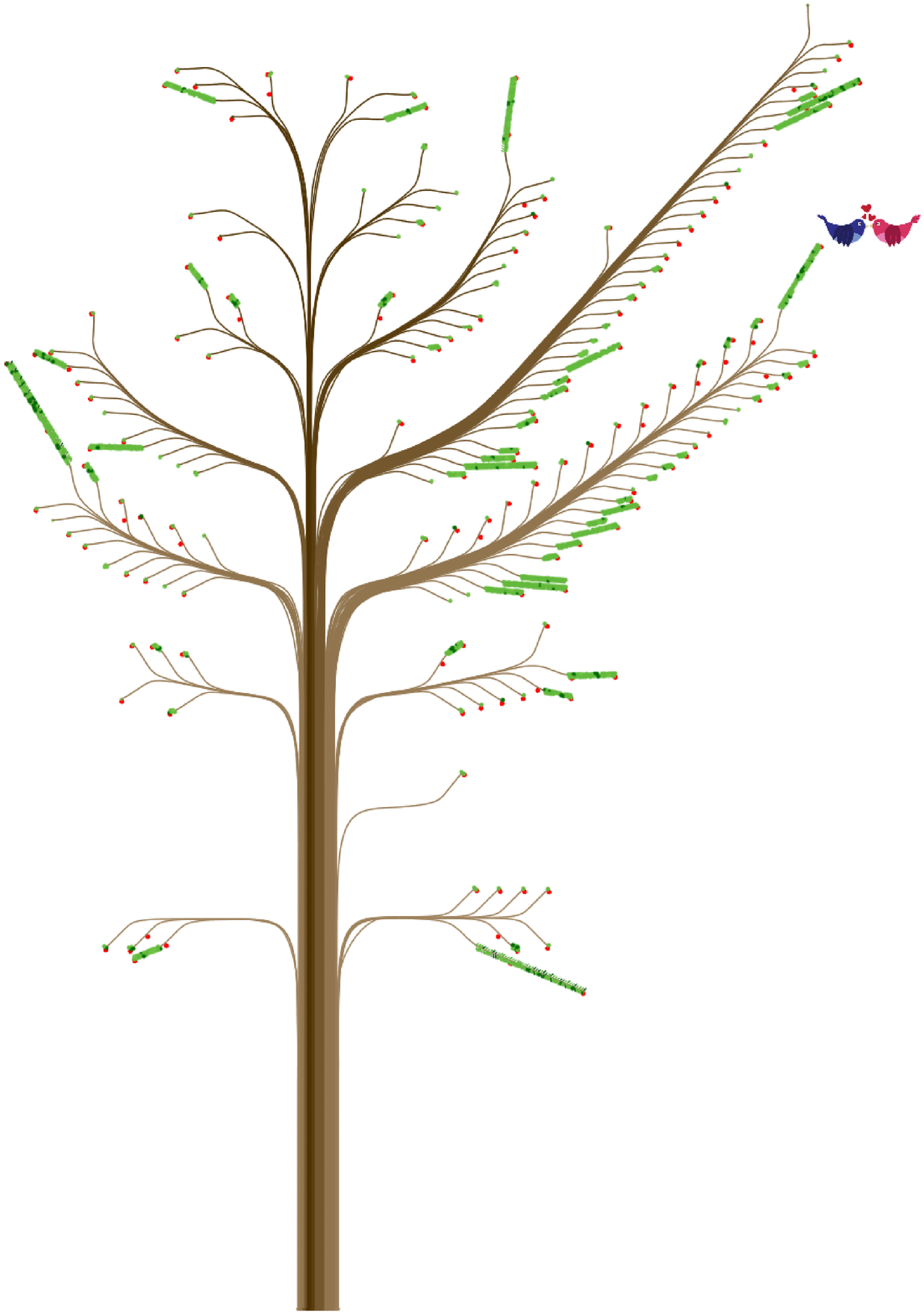}}
  \caption{The left tree represents a married man and the right one a married woman. 
They are not a married couple. 
Both have many ties of their own gender.} 
  \label{fig:married1}
\end{figure*}

Figure \ref{fig:married1} shows two trees, each having a widespread
structure. Both trees represent married persons (with
a couple of birds) having a lot of ties of their own gender
(on the same side as the birds), and one often-met tie from
the opposite sex in the same age group. Each of these
latter unique ties bears two fruits, which is the highest
value for the property ``how much the
person liked the tie." So, this particular tie is very likely
the spouse. Sometimes, these kinds of trees also
show often-met ties in the main branch of ties who are
about 30 years younger than the person (\textit{i.e.}, the second
left main branch starting from the bottom of the left tree
in Figure \ref{fig:married1}, and the first right main branch of the right
tree). These ties also have two fruits. We can reasonably
assume that they are the children of the persons.

\begin{figure*}
 \makeatletter
	\def\@captype{figure}
	\makeatother
  \centering
  \subfloat[\textit{ContactTree} of a man in 2004]{\includegraphics[width=0.34\textwidth]{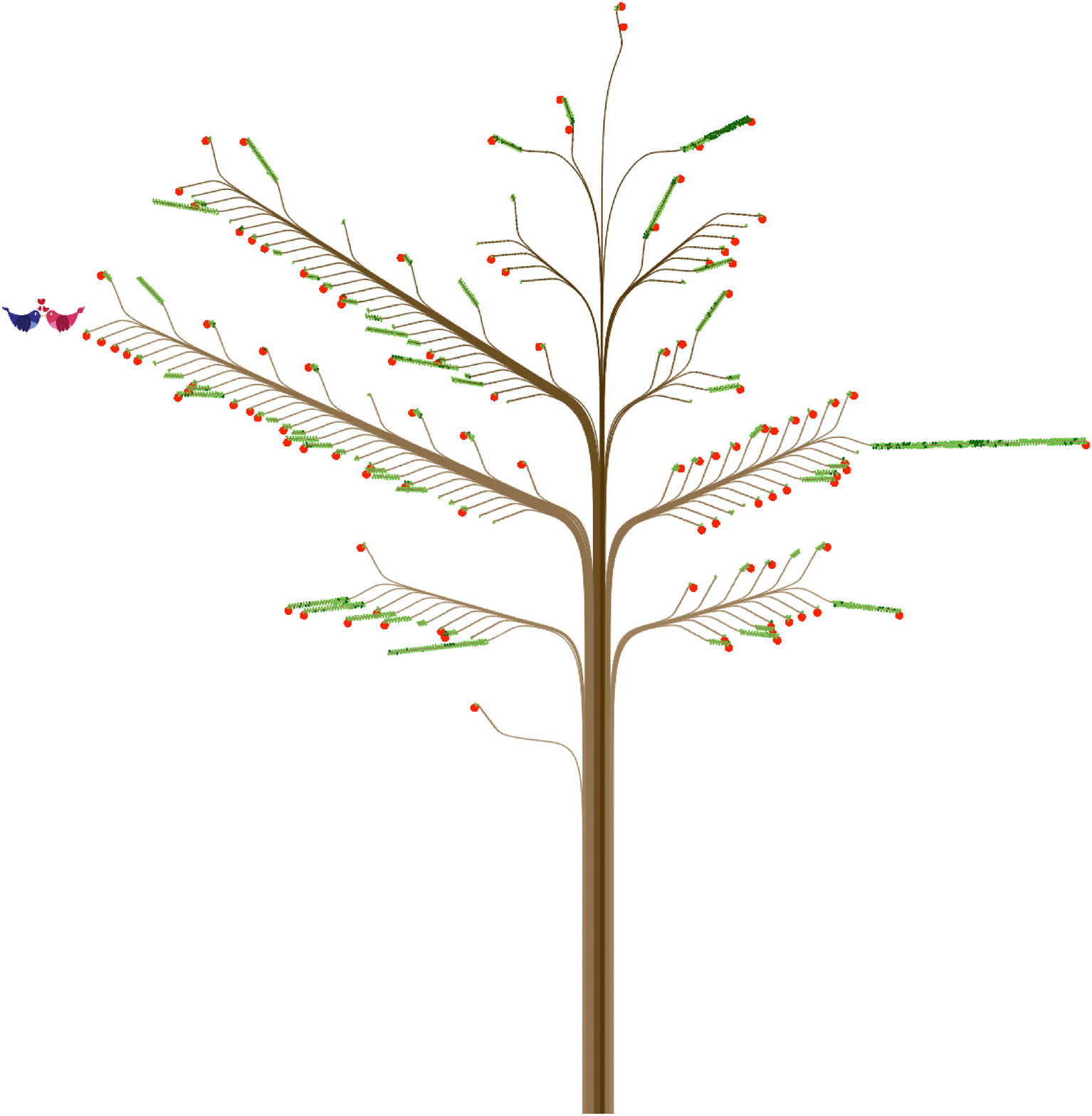}}                
  \hspace{0.01\textwidth}
  \subfloat[Same man in 2008]{\includegraphics[width=0.27\textwidth]{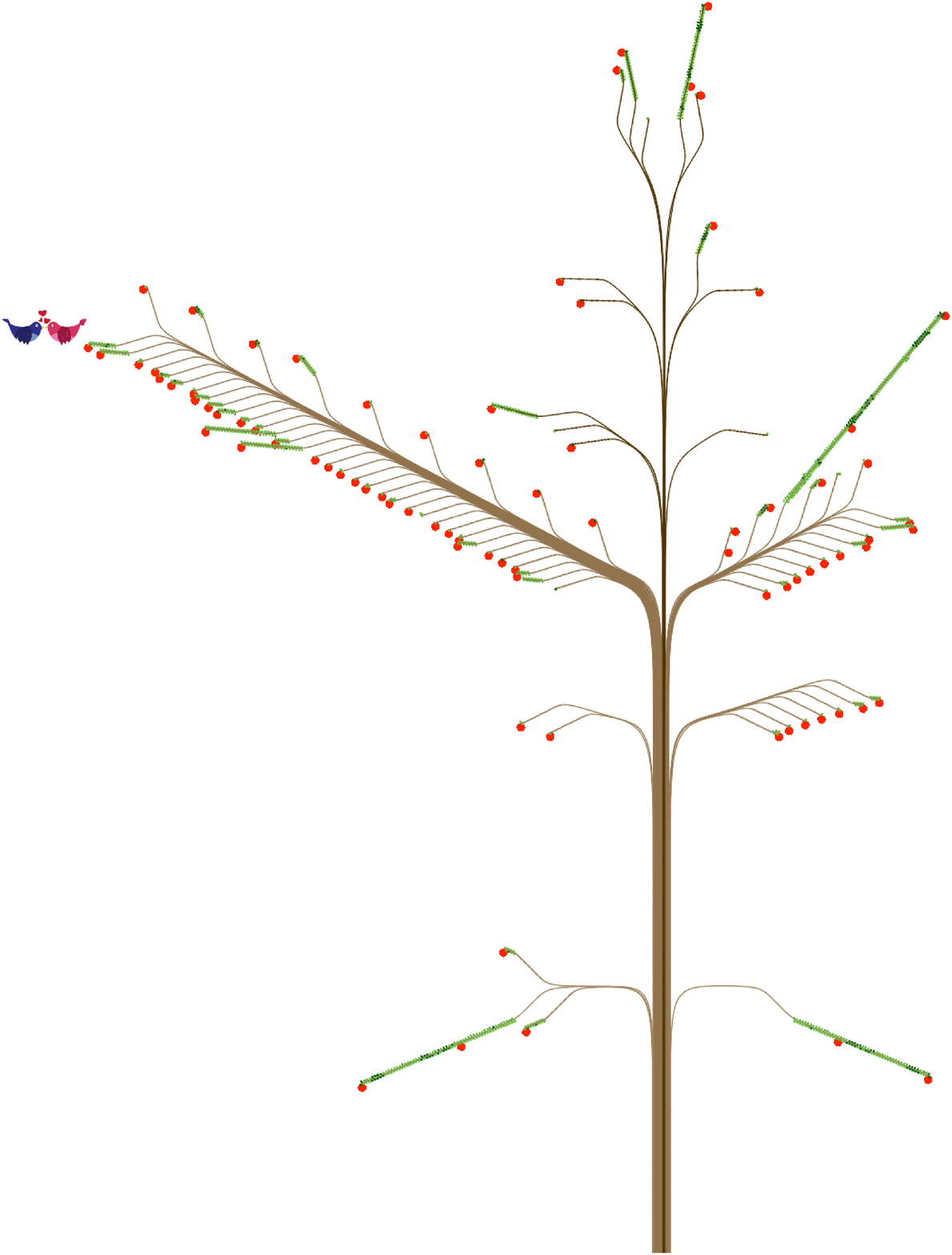}}
  \hspace{0.01\textwidth}
  \subfloat[Same man in 2012]{\includegraphics[width=0.34\textwidth]{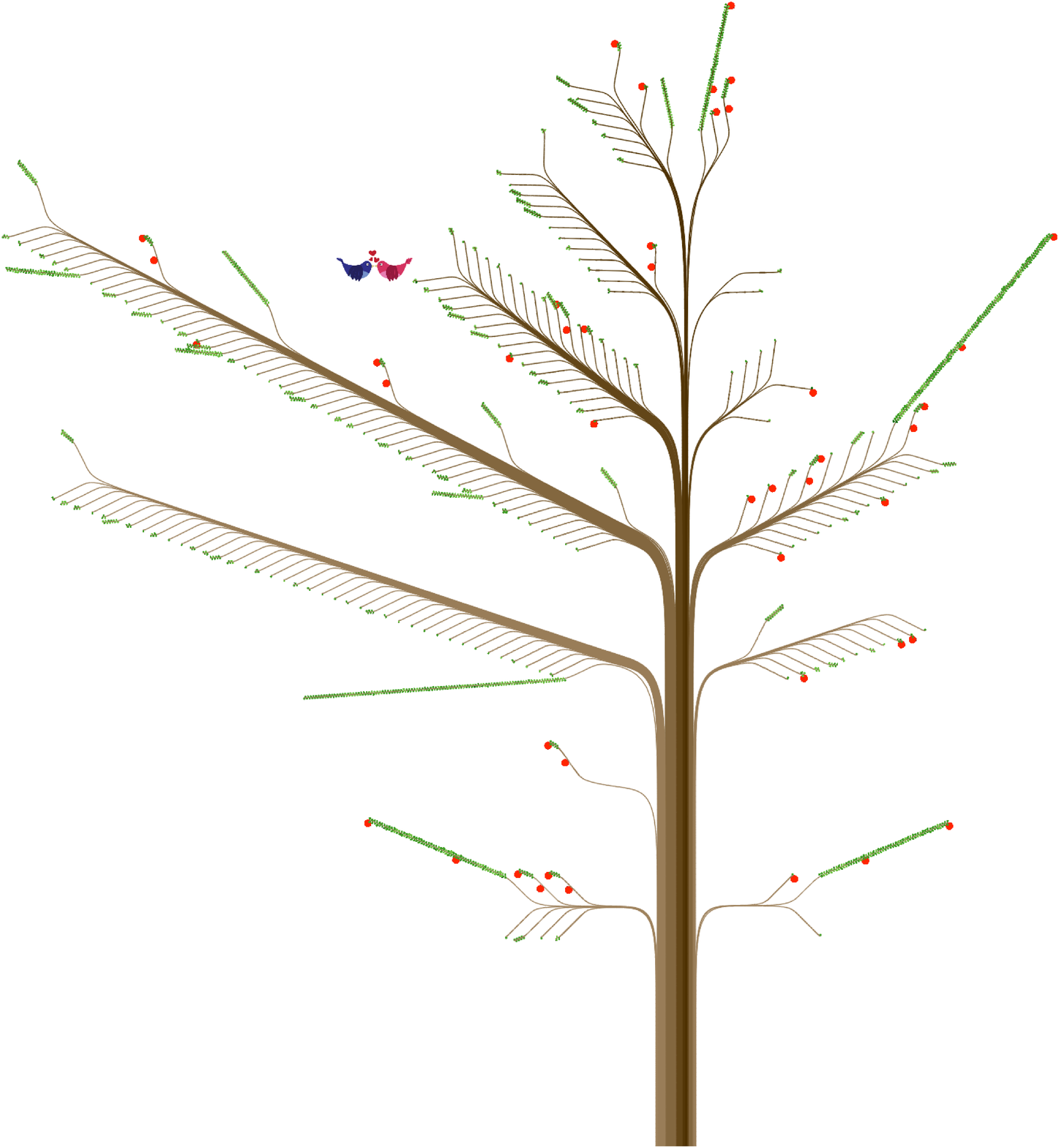}}
\\
\mbox{}\\

  \subfloat[\textit{ContactTree} of a woman in 2004]{\includegraphics[width=0.38\textwidth]{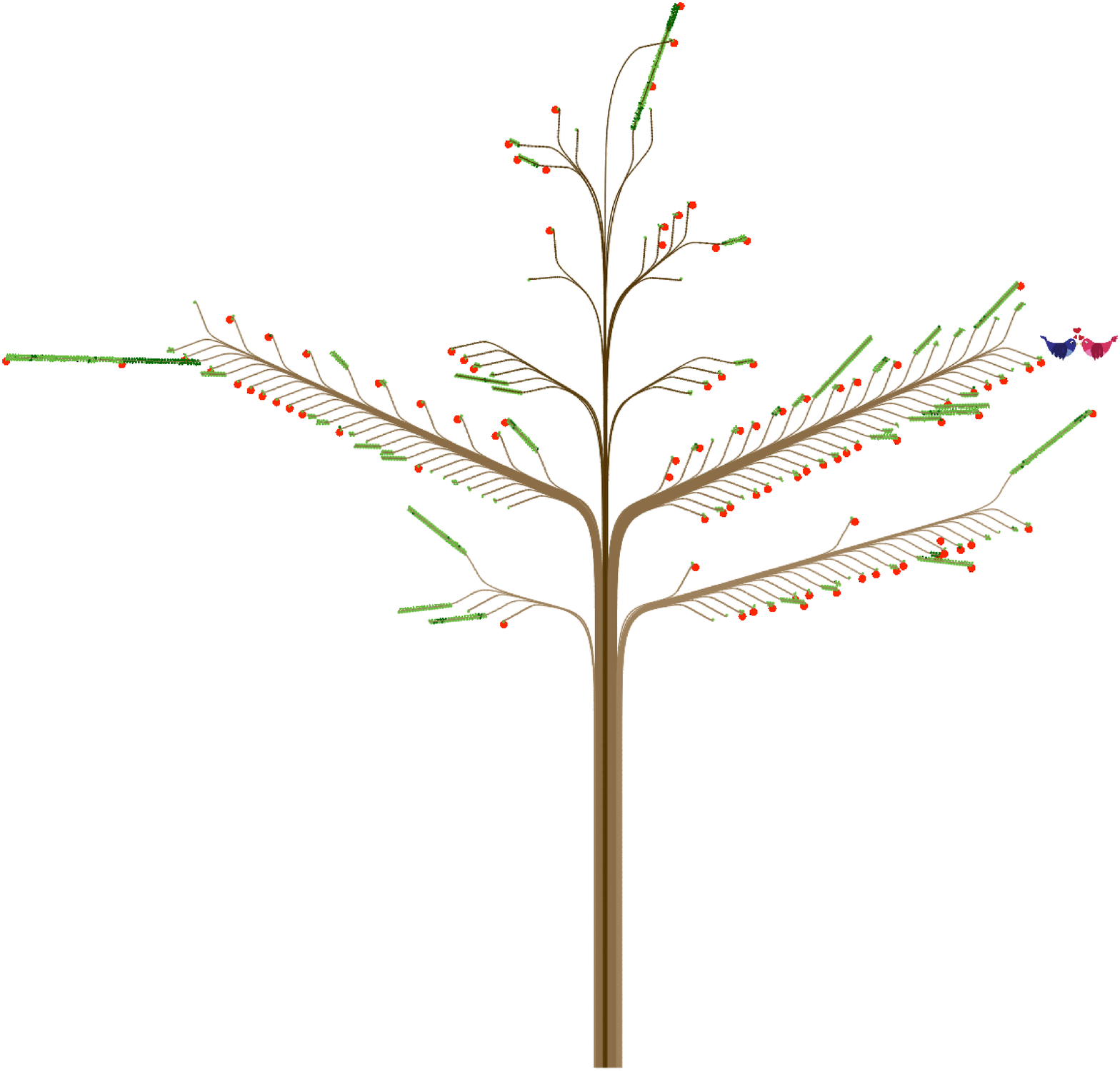}}                
  \hspace{0.05\textwidth}
  \subfloat[Same woman in 2008]{\includegraphics[width=0.20\textwidth]{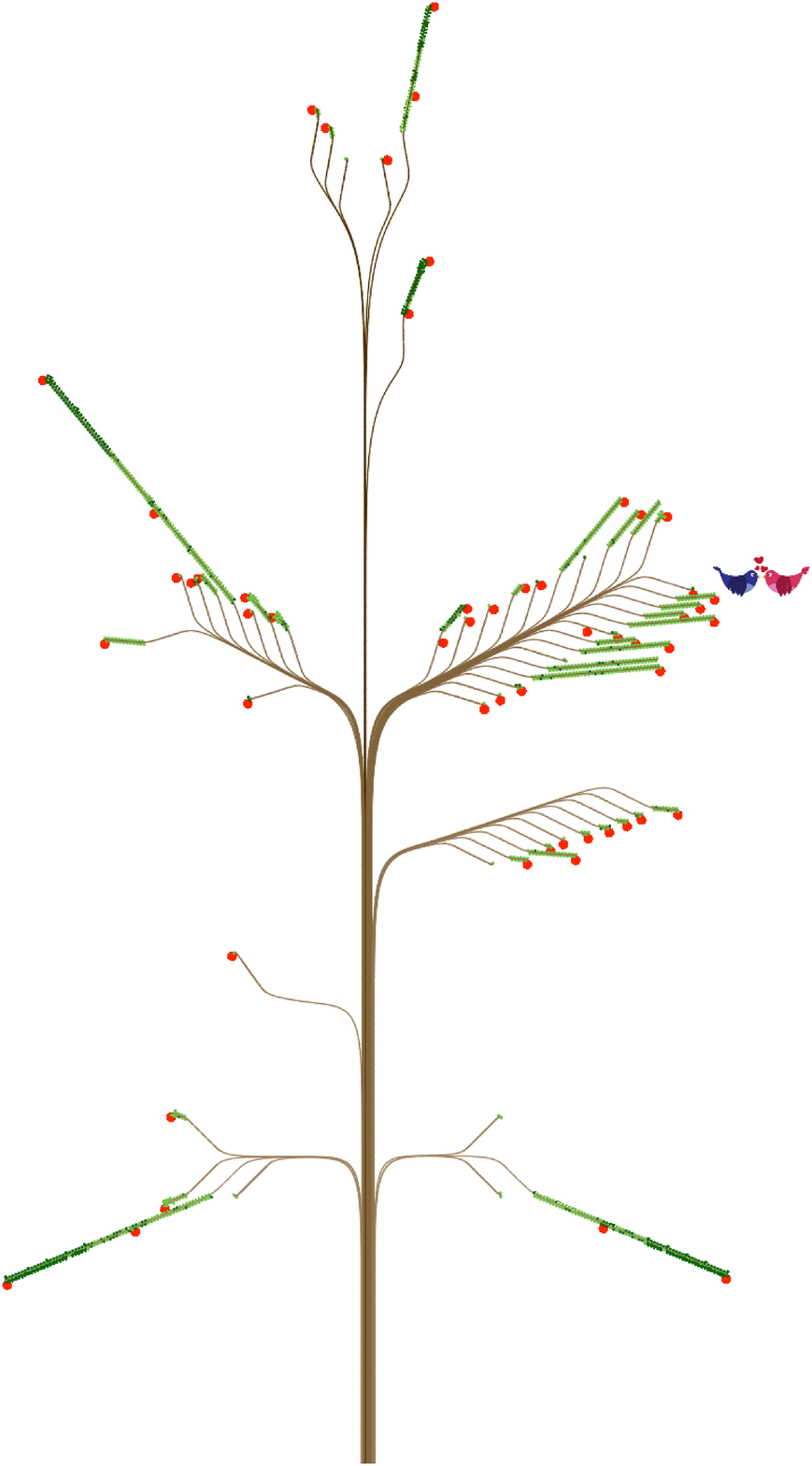}}
  \hspace{0.05\textwidth}
  \subfloat[Same woman in 2012]{\includegraphics[width=0.28\textwidth]{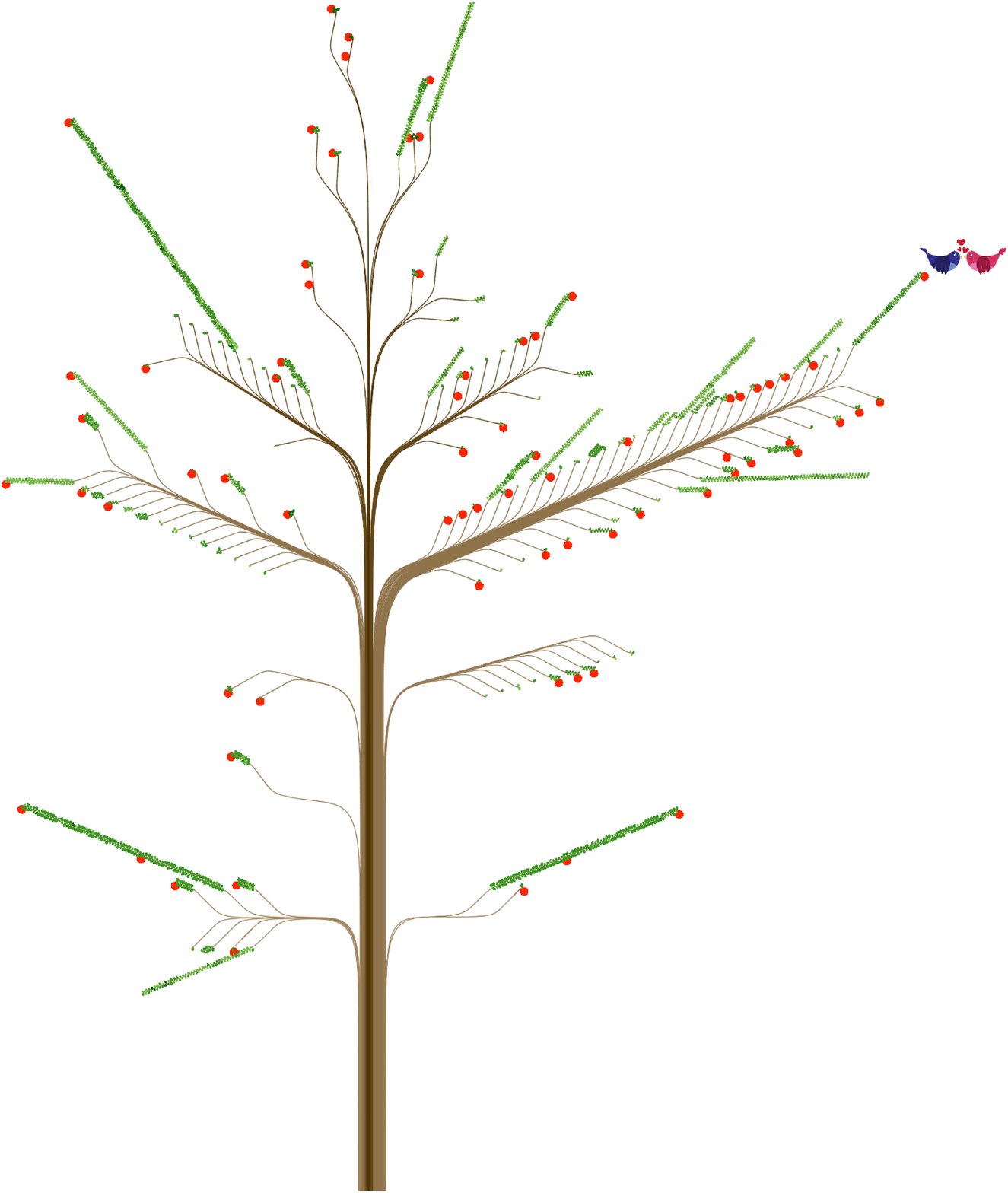}}
  \caption{Does being a parent ruin social relations? (a) and (d) represent two married persons (a man and a woman) with no children in 2004. (b) and (e) show the same persons four years later. They have two children each. The numbers of their ties and contacts have decreased. (c) and (f) show the same persons in 2012. The numbers of their ties and contacts have increased, and the tree looks pretty much the same as the one back in 2004.
  }
  \label{fig:married_parent}
\end{figure*}

Figure~\ref{fig:married_parent} shows how a married couple’s ties and contacts changed in eight years. The husband’s starting properties were similar to those of the example in Figure~\ref{fig:married1}. In 2004 the couple did not have children, and the husband’s ties and contacts were rich, particularly with those of the same gender (Figure~\ref{fig:married_parent}a). In 2008, both his ties and contacts shrunk significantly (Figure~\ref{fig:married_parent}b). Two new ties with 0-9 year olds emerged, however. These new ties were probably the couple’s children. In 2012 (Figure~\ref{fig:married_parent}c), the husband’s ties and contacts with several age groups of both genders increased, and the tree looks pretty much the same as the one back in 2004. The wife appears to experience very similar changes to those of her husband during this period: She first had abundant ties and contacts in 2004, which decreased markedly in 2008 with two new strong ties with 0-9 year olds, and then returned to a social life filled with many ties and contacts in 2012 (Figures~\ref{fig:married_parent}d, \ref{fig:married_parent}e, \ref{fig:married_parent}f). This trend suggests a correlation between the size of the overall ties and contacts and the presence of young children: Having young children seems to reduce parents’ ties and contacts with others at first, but these ties and contacts may resume as the children grow up.  

\begin{figure*}
 \makeatletter
	\def\@captype{figure}
	\makeatother
  \centering
  \subfloat[Woman having a young child in 2004]{\includegraphics[height=0.28\textwidth]{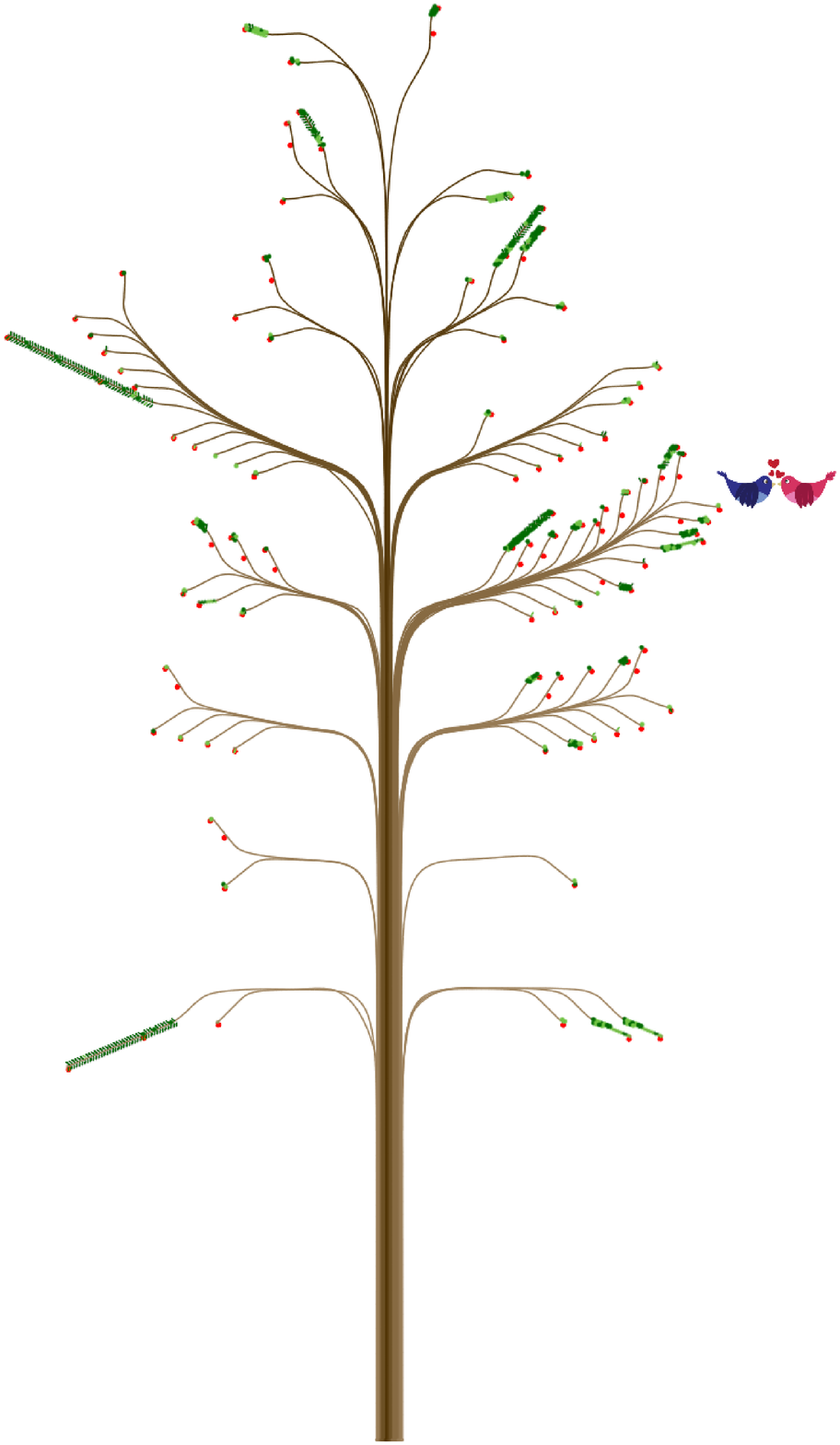}} \hspace*{0.01\textwidth}
  \subfloat[Same woman in 2008]{\includegraphics[height=0.28\textwidth]{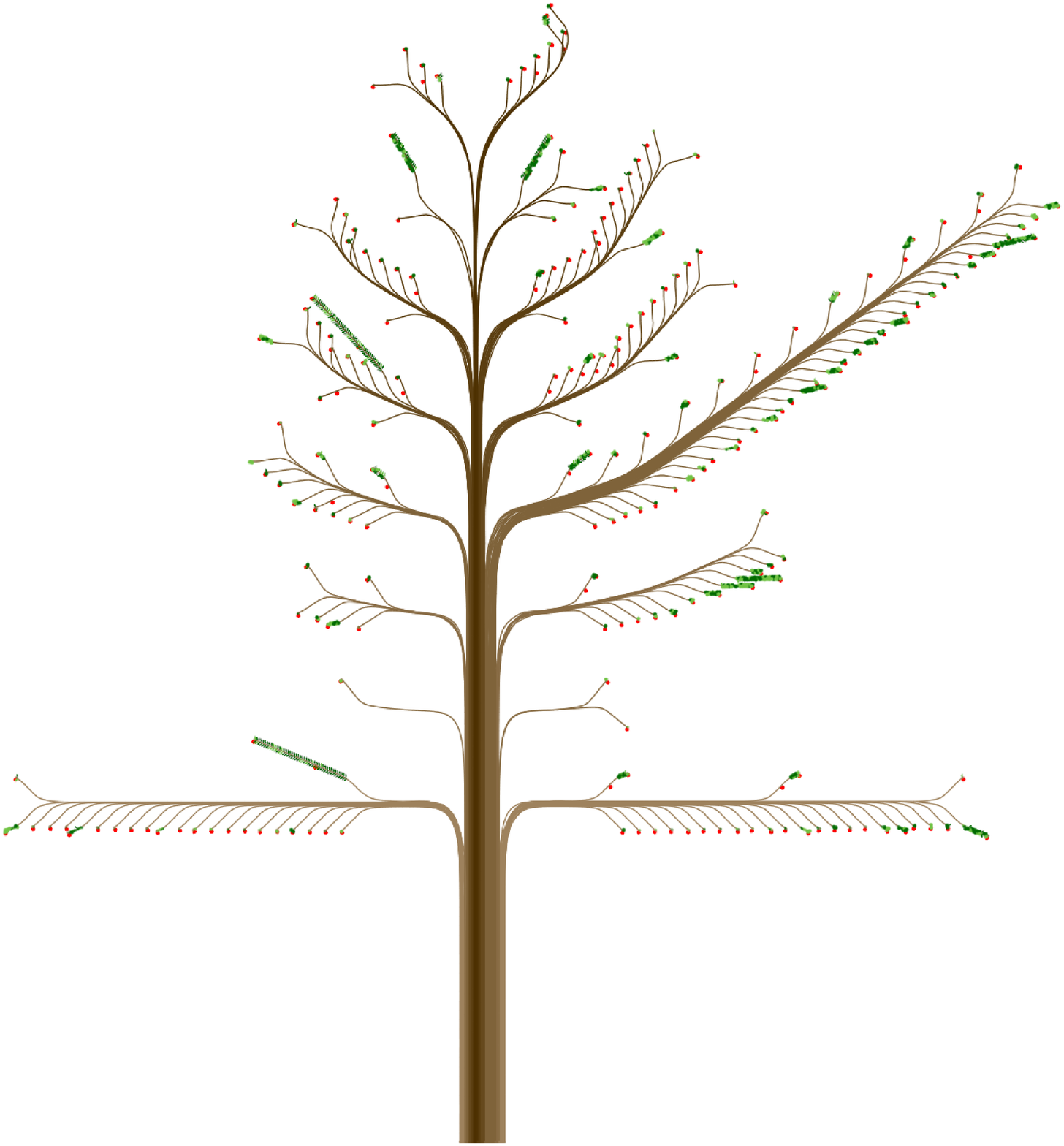}}
  \subfloat[Woman in 2004]{\includegraphics[height=0.25\textwidth]{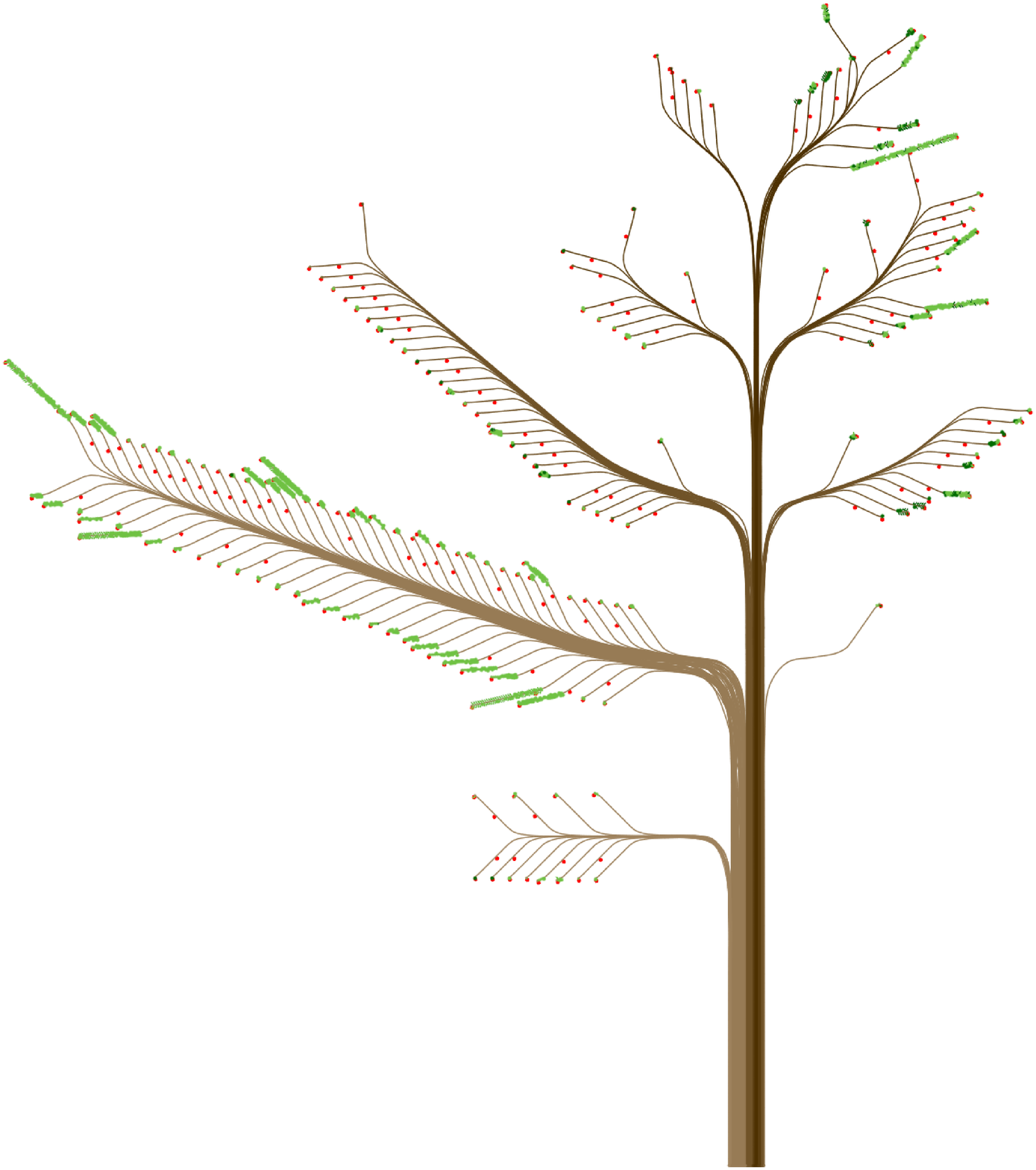}}
  \subfloat[Same woman in 2008]{\includegraphics[height=0.28\textwidth]{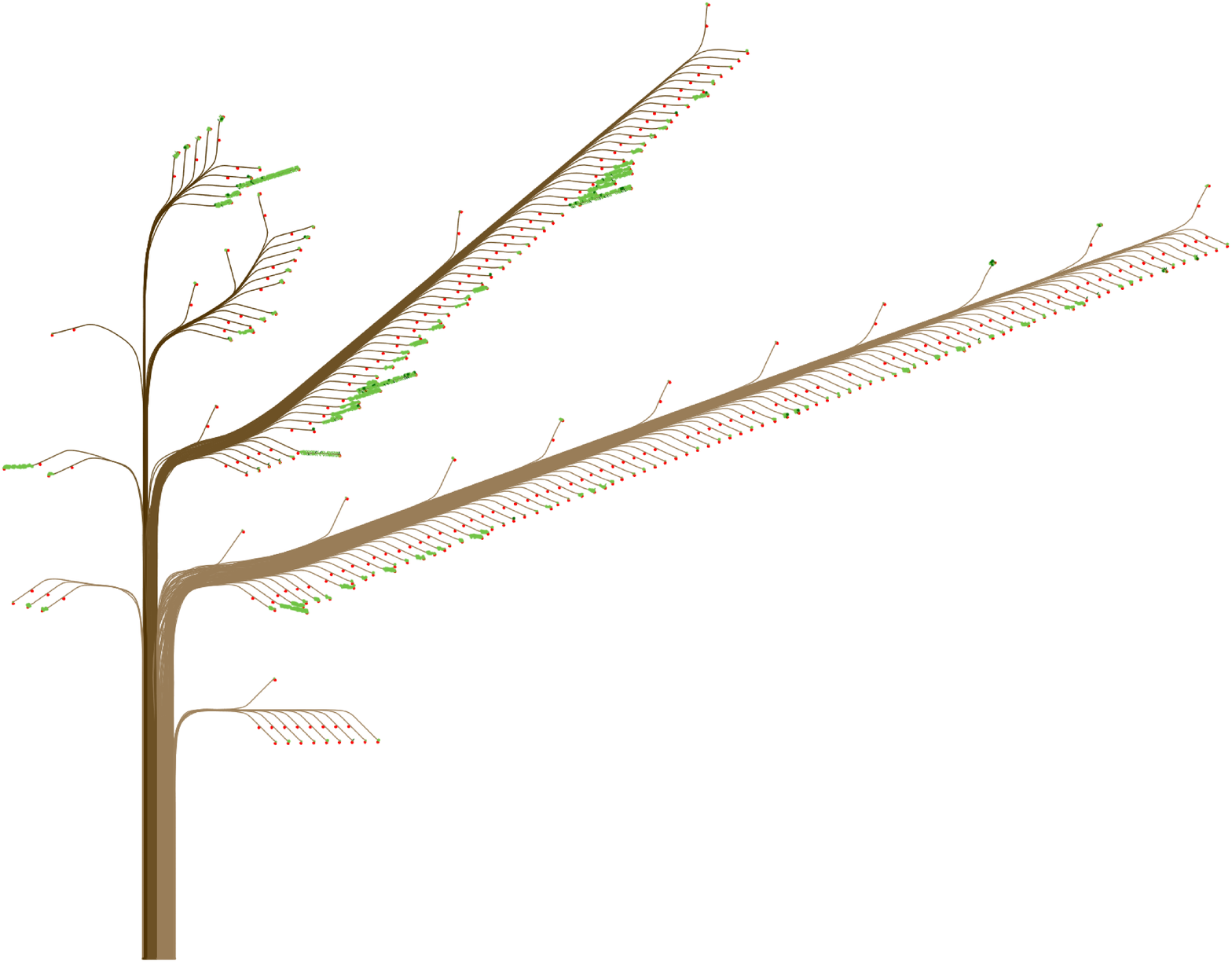}}
  \caption{Further stories from \textit{ContactTrees}.  (a) shows a woman having a young child. Four years later, as the child has grown up, the numbers of her ties and contacts have increased (b). This is likely because of her child's activities: She was meeting many children and probably their parents (especially mothers).  So the phenomenon of Figure \ref{fig:married_parent} seems not to be permanent. Figures (c) and (d) show another woman with the second mapping described in the case studies section. Most of the somewhat-liked ties of 2004 have been replaced by the very-much-liked ties in 2008. Maybe she has a new job.
  }
  \label{fig:married_conclusion}
\end{figure*}

Such fluctuation along one’s life course is not uncommon. In Figure~\ref{fig:married_conclusion}, for example, the person represented is a married woman who had a child and did not have a large number of ties and contacts in 2004, a pattern consistent with our last example. We know that the child was between 1 and 4 years old in 2004, because he was represented on the lowest main branch of the left side of the trunk (Figure~\ref{fig:married_conclusion}a). Four years later (in 2008, Figure~\ref{fig:married_conclusion}b), the woman’s ties and contacts expanded significantly. This change is likely because of the child’s activities: She was meeting a lot of his friends and probably their parents too (especially mothers). Thus, we can speculate that the decrease of the number of ties and contacts when one has a young child may not persist, as exemplified by the previous case of the married couple.

\subsection{Second mapping}\label{second_mappin_cd}

Also based on the \textit{Contact Diaries} dataset, this second method deals with another mapping. With this mapping, the person who is represented by  a \textit{ContactTree} likes very much those ties lying on the right side of the trunk, and  likes  somewhat those ties on  the  left  side. The  position of a main branch along the trunk indicates how many years the person has known the ties of this branch. The lowest main branches (on both right and left sides of the trunk) are strangers, the second lowest branches hold ties known for less than 1 year, the third stand for those ties known for 1 to 4 years, the fourth are between 5 and 19 years, and the upper main branches are ties known for at least 20 years. The small branches growing above (below) main branches represent ties who are more (less) than 40 years old. A branch with one (two) fruit(s) is a male (female) tie. Mapping of leaves is still the same: Each leaf represents a contact, its size is proportional to the length of the contact, and its darkness indicates how positive or negative the person felt after the contact.

\begin{figure*}
 \makeatletter
	\def\@captype{figure}
	\makeatother
  \centering
  \subfloat[This woman has 819
ties, with 4,091 contacts. Her \textit{ContactTree} is the
largest in the \textit{contact diaries} dataset.]{\label{fig:ego46}\includegraphics[height=0.35\textwidth]{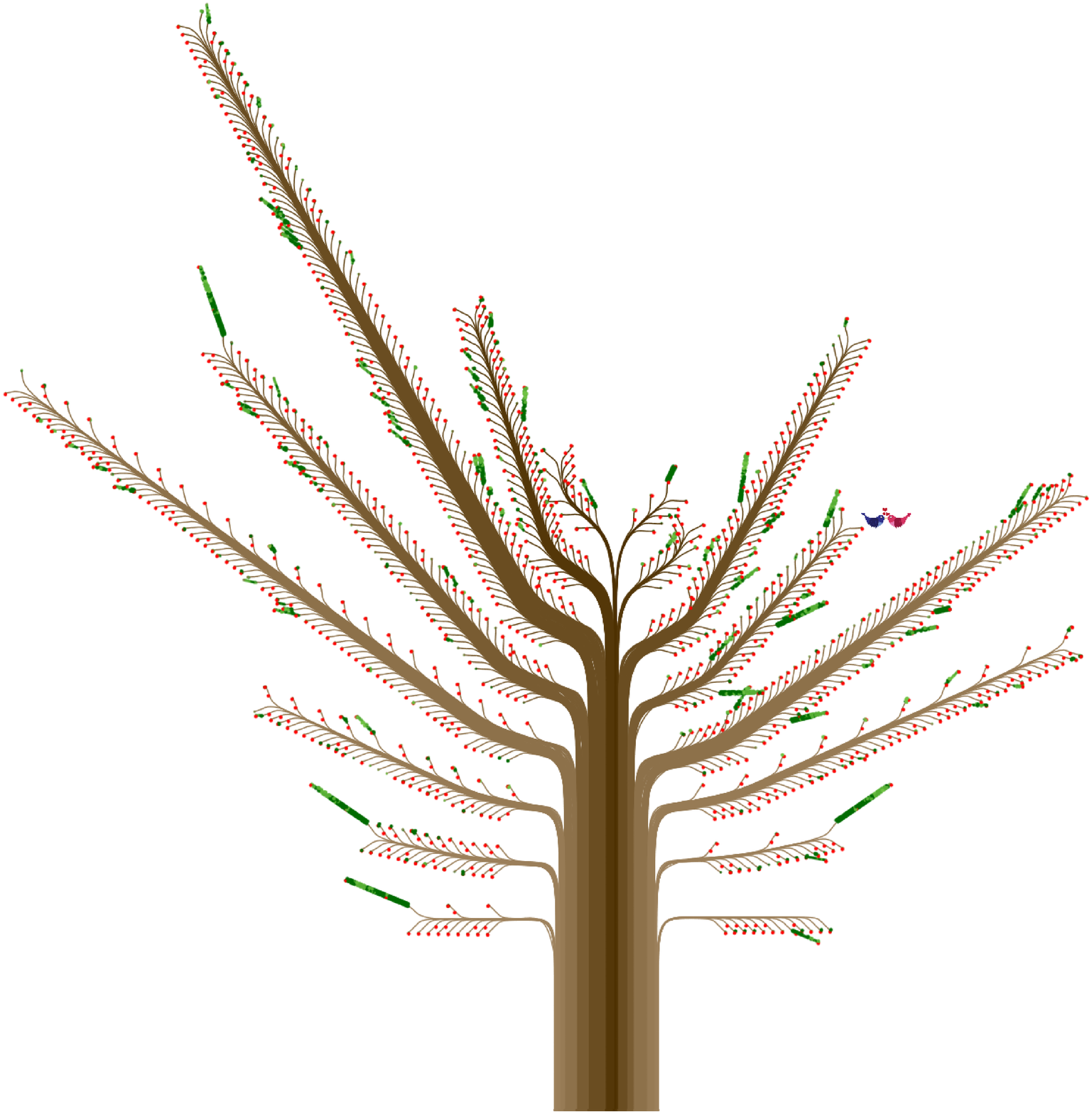}}
  \hspace{2cm}
  \subfloat[Two small branches intersect]{\label{fig:cross}\includegraphics[height=0.27\textwidth]{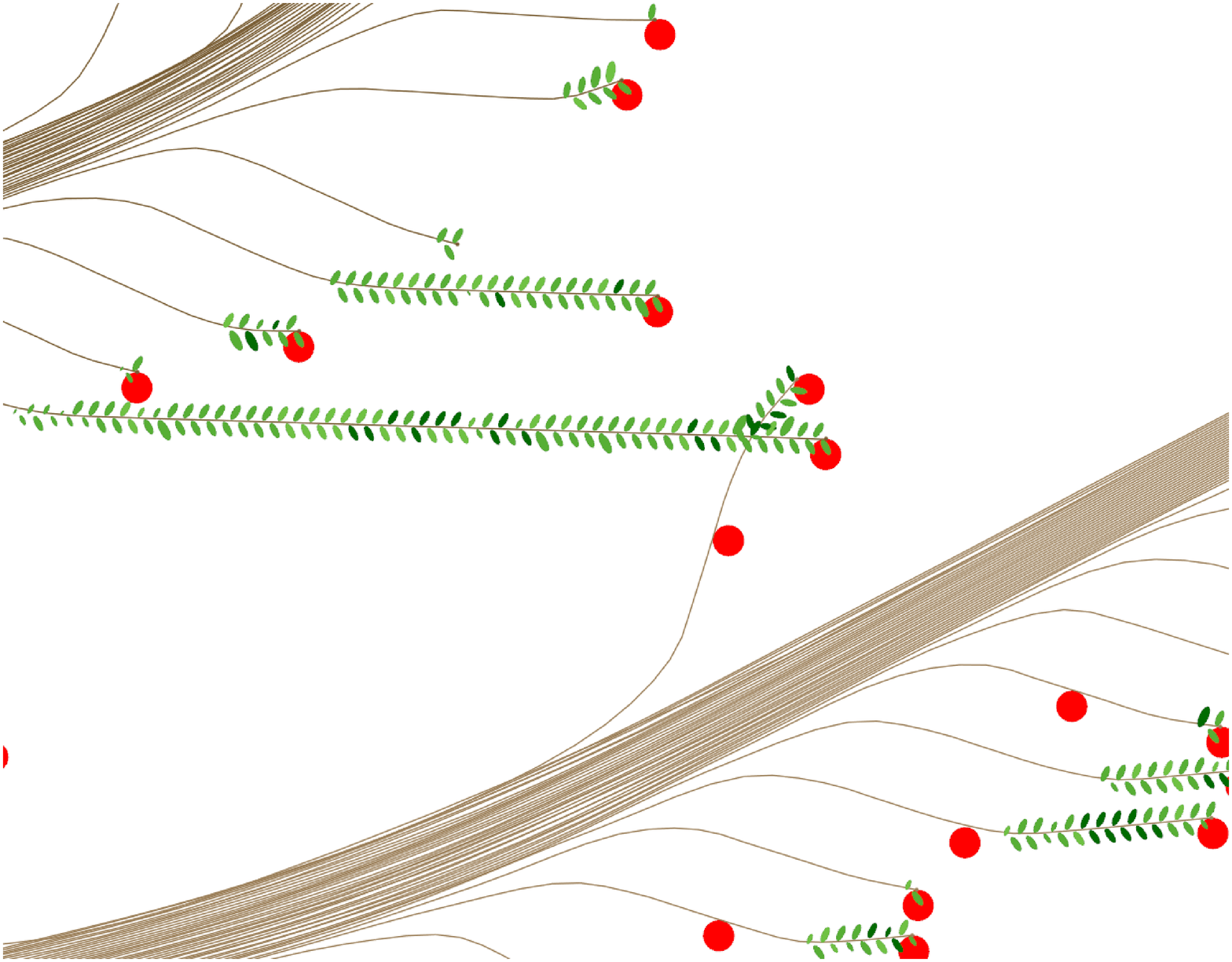}}
  \caption{Limitations. 
  }
  \label{fig:change_job}
\end{figure*}

Figure \ref{fig:change_job} shows two \textit{ContactTrees} based on this mapping. They represent the ties of the same young woman in 2004 (when she was 25) and in 2008. Most of her ties had been known for less than 5 years (the first, second, and third main branches starting from the bottom). Both \textit{ContactTrees} are very unbalanced, but not on the same side: The woman was less enthusiastic about most of her ties in 2004, while she liked  her ties in 2008 very much. We can also see that her ties in 2004 were well balanced in gender and age (fruits and side of main branches). In contrast, in 2008, most of them were women under 40 years old. We cannot reasonably think that the ties remained the same after four years and that the person had changed her mind toward them because of the difference of their gender/age. So she had not kept her ties of 2004 and had met a lot of people between 2004 and 2008. In compliance with these observations and the large number of ties for each period, the most reasonable hypothesis is that she had left her previous professional position and had gotten a new job.

\subsection{Observations}

Different mapping methods of our \textit{ContactTrees} aim to highlight various properties of social relations. Inspired by the botanical metaphor, we incorporate key features of a tree structure, such as length, location, shape, size, and color into our design. Applying these methods to two data sets, we have been able to distinguish how ties and contacts/collaborations vary by comparing main branches, small branches, leaves, and fruits. Some of the observations we made from these trees can be further verified by other empirical data sources. For example, in the contact  diaries dataset, we made several inferences about a person's overall patterns of ties and contacts with others, and highlighted some trends of social  relations  throughout his or her life span. When we dig deeper into the original data, and even re-interview those persons whose \textit{contact diaries} were used in our illustrations, some of these observations and hypotheses can be confirmed or modified.

For example, the small branch in Figure \ref{fig:married1}a that bears
the most leaves on a main branch parallel to the one with
two birds on the opposite side (left) of the trunk is
indeed the focal person's spouse. The same is true
about what we learned from Figure \ref{fig:married1}b. Likewise, the
small branches with the most leaves on the first and second
bottom main branches in Figure \ref{fig:married1} are these focal persons'
children.

Our readings about how trees change over time also
turn out to be very close to the actual social life of the woman
shown in Figure \ref{fig:married_conclusion}. When her 2-year-old son stayed
home all day in 2004, she was actually pretty active
by keeping contact with relatives, neighbours, friends,
and strangers. Among her 455 ties, about 77\% had no
specific relationship with her. Four years later  when
her son attended the kindergarten, she started to build
many new ties and made new contacts with the
boys  schoolmates  and  their  parents  (the  indirect  ties
via her son grew from 4 to 114, and the contacts out
of such indirect ties increased from 13 to 436). Equally
important, she still maintained contact with ties in
nearly all other categories, even strangers. As a result,
the number of her ties increased by 28\% (to 581), while
the number of contacts went up by 42\% (from 1,136 to
1,618). Compared to the increase in ties, the magnitude
of the number of contacts may have contributed
more to the visualization effect in the second (larger) tree
in Figure \ref{fig:married_conclusion}.

When we compare social relations over time in other
cases, the trends may become too complex to  fully
understand just by looking at how trees (by a single
design) change in two different years. For example, the
two pairs of trees in Figure \ref{fig:married_parent} 
turn out to be husband
and wife. In 2004, when the husband (an engineer with
an environmental protection firm) was 33 and the wife
(a specialist in a tele-communications firm) was 32, the
couple was newly wedded and childless. Back then, they
each had quite a few ties and contacts with co-workers,
friends, weakly tied acquaintances, and even strangers.
When they had 2 young children between 2004 and 2008,
both of them indeed had to cut down some of their former
ties  because of  the  new  family obligations.  
The husband had moved to a highly bureaucratic
optronics firm, however, where he spent most work hours with
machines and a limited number of colleagues; These new job 
conditions limited his ties and contacts with others.

While the couple spent more time with their young children, the wife also cut down her ties with others. She continued to work in the same firm with the same position. Even though her ties decreased from 361 in 2004 to 101 in 2008, the total number of her contacts only dropped from 1,522 to 1,434. The significant decrease in ties is clearly visible by comparing the branches in Figures \ref{fig:married_parent}c and \ref{fig:married_parent}d. In contrast, the slight change in the number of contacts may have been overshadowed by the sharply decreased number of ties. In particular, she retained as many contacts with her colleagues, relatives, and weakly tied acquaintances over the 4 years, even though these contacts were limited to a much smaller number of persons.

Such changes may lead to different implications about our mapping strategies, botanical metaphor, and substantive issues in social networks. For example, the first mapping on the \textit{contact diaries} dataset may be better at catching the overall structure of gender, age, and relations. In contrast, the second mapping may be more sensitive to other properties, especially the strength of ties with contacted persons. Further, as our visualizations clearly illustrate, a tree's general visual effects may lie more in the overall structure of the trunk, main branches, and small branches, rather than in the number of leaves. Such a tendency of visualization effects may coincide with the overall stress on interpersonal ties in  the social network literature. It seems more direct to demonstrate and assess the conditions of a social network by its total size and overall structure. As discussed with regard to the change in Figure \ref{fig:married_conclusion}, however, other properties (especially contacts) may add equally critical information to that structure. Therefore, while branches make up the main structure of the tree, the leaves and fruits give ties' details while making the tree more natural, which is fundamental and essential to our botanical metaphor.


\section{Discussions} \label{sec:discussion}

The potential effectiveness of our visualization can be evaluated from two perspectives. First, there could be limitations to the algorithm itself and its scalability. Second, our method differs markedly from the previous related network visualizations in its capacities of dealing with a more sophisticated egocentric network beyond the usual structure of nodes and edges.

\subsection{Limitations of the algorithm}

Figure \ref{fig:ego46} shows the woman who has the highest numbers of ties and contacts in the \textit{contact diaries} dataset. As one can see, the representation is less aesthetically pleasant than the previous trees, even though our technique fulfills its purpose: The tree shape is still correct, botanical tree-like, and each feature can be easily distinguished.

Highly unbalanced \textit{ContactTrees} also lead to less aesthetically pleasing representations. Figure \ref{fig:change_job}.b shows an example. The woman likes very much many of her ties, which are positioned on the right side of the tree, as indicated by two long main branches. These long branches make the tree less realistic, but it does not disrupt the readability of the data.

The only concrete problem we observed in our datasets concerns branches with many leaves. Sometimes these branches intersect. Figure \ref{fig:cross} shows an example of this phenomenon. This kind of occlusion problem is not frequent and does not mislead the user about the global shape of a tree, a main branch, or a small branch, but this can be annoying when one focuses on a specific contact. Increasing the gap between consecutive main branches would eliminate this problem, although it also increases the free space between tree elements.

\subsection{ContactTrees vs. alternative visualizations}
\label{compare}

Among tools that help visualize social networks, \textit{Vizster} is powerful and well-cited \cite{Heer_2005}. Based on the visualization of a node-link diagram representing individuals and ``friendship'' relations, it offers many interaction techniques that help select individuals and ties one wants to visualize or highlight.
\textit{Vizster} shows how ties of several individuals are 
related among a large social network, and helps discover connections between individuals, paths and community structures. That is, it performs tasks specific to visual graph analytics.

The problems addressed in this paper are somewhat different. First, we deal with an individual's ties with his or her network members, but not the ties among these members. As with the case for many other egocentric networks, 
we do not know if a network member who is connected to the focal person is also connected to another member. Then, 
a visualization better suited for complete networks is less compatible in  our case. 

Second, the \textit{contact diary} dataset includes some information about individual's ties that cannot be readily displayed or identified in a tool like \textit{Vizster}. For example, it would take extraordinary efforts to readjust the tool to show gender and age distributions, as well as who are more favored by the focal person in the same graph. It would be even more challenging to display how these properties evolve over time for a given person. While a tool like \textit{Vizster} tends to focus more on showing the whole network, our design can neatly organize and highlight the attributes of each specific ties. 

Third, the \textit{contact diary} dataset contains various attributes about each unique contact between a focal person and the persons he or she encounters. These attributes at the contact level cannot be incorporated into a visualization tool such as \textit{Vizster}. In comparison, our design takes into account the attributes at both tie and contact levels, thus disclosing the eatra and rich information embedded in a sophisticated data structure. In sum, \textit{Vizster} and \textit{ConatctTrees} are based on two paradigms that do not fill the same tasks and so can be complementary but not concurrent.

A similar botanical visualization tool was proposed \cite{Kleiberg_2001} to reveal huge hierarchies. The main difference between this tool and our \textit{ContactTrees} lies in the nature of the input data. Their purpose is to visualize a hierarchy while ours is to visualize a more complex and specific data structure of social networks. Applying their visualization to this structure is not trivial. We would need to find a system for extracting a hierarchy according to ties' and contacts' attributes, and we are convinced that designing a specific visualization is more efficient. Moreover, their visualization is a 3D design and thus induces occlusion issues. The technique they use to generate their tree could also help us construct a tree following the same requirements as ours. However, we decided to use bundled curves instead of cones for several reasons related to the selection of ties (see the video for an example).

\textit{TreeVersity} \cite{Gomez_2012} is an interactive visualization that allows users to detect node value changes and topological differences between trees. It has been developed to deal with hierarchies and so includes interaction features specific for this purpose. A \textit{ContactTree} does not represent a hierarchy. There are not internal nodes and the leaves are not unique for each branch. Moreover, our application dataset doesn't hold information such that a given leaf in a 2004 \textit{ContactTree} corresponds to a particular leaf in a 2008 \textit{ContactTree}, which makes most of the \textit{TreeVersity} comparison techniques useless for our purpose.


\section{Conclusion and future work} \label{sec:conclusion}

\textit{ContactTrees} enables one to map ties and contacts based on various priorities and preferences. Facing overly sophisticated network  data,  whether  they are taken from \textit{contact diaries}, citation records, survey archives, or online social media, researchers and users would benefit greatly by first looking at the key features of a tree. These features capture not only ``ties," the underlying unit of analysis in most social network analyses, but also ``contacts" among ties. Without contacts, we would miss the key information that is underlying all social relations. Without leaves and fruits, in the botanical terms, we could hardly judge a tree’s main characteristics or how it grows over seasonal changes and other life cycles. By better capturing the structure and dynamics of social relations and interpersonal interactions, which in turn facilitate instant comparisons of how network structures differ, the addition of this key property thus distinguishes our design from the majority of other visualization tools for social networks. In other words, furnishing leaves and fruits to the trunk and branches greatly enhances the capacities of visualization tools.

Several shortcoming remain to be refined and revised. For
example, more mapping based on other principles may be needed to further illustrate or clarify how ties and contacts
change over time. Because the quality of ties plays a
significant role in the overall health of an egocentric network, future
work may also want to elaborate how to better display
the varying conditions in trunk and branches. In addition, the
case studies we use to map our \textit{ContactTrees} were taken
from a relatively small number of individuals in a single society. 
When compatible samples become available from other societies or cultures, the design can be applied and further tested for more universal usefulness.

The actual use of \textit{ContactTrees} can be supplemented with other formats or sources
of network data.  Among  the  abundant  resources  about
social interactions, the booming online social media 
represent one  of  the  most  challenging  yet  promising
sources upon which  \textit{ContactTrees} can be applied with
ease for both researchers and other end-users. Our design 
can also be extended to other domains to help identify 
patterns and trends of social interactions. In particular, 
alternative formats of \textit{ContactTrees} could show 
overall ties and contacts for a specific age
cohort, occupational group, or the whole population 
within a geographical area. In that case, our design 
would facilitate knowledge discovery on social networks 
at the micro- and macro-levels alike.

\section*{Acknowledgments}

This research is sponsored in part by the U.S. National Science Foundation and UC Davis RISE program. 

\bibliographystyle{pnas}
\bibliography{biblio}









\end{document}